# NOE-less protein structures-an alternative paradigm in Highly Paramagnetic Systems

Inês Trindade[§], Michele Invernici[£], Francesca Cantini[£], Ricardo Louro[§]*, Mario Piccioli[£]*

[§] Instituto de Tecnologia Química e Biológica António Xavier (ITQB-NOVA), Universidade Nova de Lisboa, Av. da República (EAN), 2780-157 Oeiras, Portugal, louro@itqb.unl.pt.

[£] Magnetic Resonance Center and Department of Chemistry, University of Florence, Via L. Sacconi 6 50019 Sesto Fiorentino, Italy.

**ABSTRACT:** The current paradigm for protein structure determination by NMR spectroscopy has thus far been based on collecting distance restrains between pairs of nuclei in the form of nuclear Overhauser enhancements (NOE). These, have in some circumstances been supplemented with other information sources such as paramagnetic relaxation enhancements or residual dipolar couplings. Here, we report for the first time a protein structure determination by NMR without the use of NOEs. The protein PioC from Rhodopseudomonas palustris TIE-1 is an HIPIP where the 4Fe4S cluster is paramagnetic and provides the source of Paramagnetic Relaxation Enhancements (PRE) used as alternative distance constraints. Comparison of the family of structures obtained by NOE structural restraints, with that obtained by PRE and with the family of structures obtained by combining NOEs and PREs reveals that the pairwise RMSD between them are similar and comparable with the precision within each family. This work sets the stage for the structural characterization of small and dynamic paramagnetic metalloproteins opening a new paradigm in the use of NMR in structural biology.

Metalloproteins represent 40 to 47% of all known enzymes[1,2] and, for all of them, the metal center(s) are essential for catalysis, electron transfer, metal storage/transport, or they play a crucial role in stability and structural properties[3-8]. Structural biologists are mainly interested in obtaining detailed information in the proximity of the metal center(s), where the biochemically relevant events occur. NMR is a privileged approach for characterizing metalloproteins, since it can provide the structure in solution at atomic resolution, amplitude and time-scale of local dynamics[9,10] and hints on the electronic structure and oxidation states of the metal center[11], in conditions that mimic the physiological context. However, a significant part of the metalloproteome contains a paramagnetic metal ion that enhances nuclear relaxation in its vicinity posing a challenge for signal detection[12]. In consequence, NMR solution structures of paramagnetic macromolecules have required a combination of classical structural constraints (NOE, residual dipolar couplings and secondary structure predictions) and paramagnetism-based constraints (cross-correlation, pseudocontact shifts and paramagnetic relaxation enhancements) [13-16]. High resolution NMR structures were obtained for paramagnetic metalloproteins [17-21] and for proteins containing paramagnetic metal binding tags [22-26]. The substitution of the paramagnetic center with a diamagnetic analogue has also been used to obtain NMR structures of metalloproteins[27,28]. A number of alternative approaches have been proposed, in which sparse NOE and chemical shifts based structure prediction methods succeeded to obtain well-defined solution structures [29,30]. Indeed, structures without NOEs have been solved when accurate orientational constraints are available from at least two full sets of rdc[31,32]. However, NOEs still constitute the fundamental building block to sample the protein conformational space and provide a unique fold [33-35]. Indeed, networked interatomic distances are more powerful and robust than orientation constraints and dihedral angles. In the case of $Ln^{3+}$-substituted Calbindin $D_{9k}$, it was found that even in the presence of a very large number of rdc and pcs a minimal number of about 30 NOEs was necessary to obtain a structure[36]. Cross correlation rates, and especially that originating from the interaction between Curie-spin and dipolar coupling, may also contribute to solution structure calculation[37,38]; in combination with orientational constraints measured on a diamagnetic analog, they succeeded to obtain a structure without NOEs [39]. PREs are dipole-dipole restraints, like NOEs. Therefore, if a sufficient number of them is available and if they can be measured throughout the entire protein, they should sample the protein conformational space with an efficiency comparable to NOEs, even if they are a set of distance constraints all involving a single point, i.e. the paramagnetic center. An important drawback of PREs is that NMR experiments currently available to measure $^1$H $R_1$ and $R_2$ rates [40] fail to provide PRE values in the proximity of the paramagnetic center, where many signals are broadened beyond detection. This can be overcome by developing experiments capable to measure with good precision fast relaxation rates. For longitudinal relaxation, the IR-HSQC-AP experiment [41] is very effective to identify and to measure $R_1$ rates of fast relaxing signals. However, it has been reported that $^1$H $R_1$ values are more susceptible than $R_2$ to internal motions and cross relaxation[42]. Therefore, accurate measurements of both rates are important to obtain reliable information on the metal-to-proton distances.

To explore this issue, the NMR solution structure of the small Iron-Sulfur protein PioC from *Rhodopseudomonas palustris* TIE-1 [43] represents a paradigmatic challenge. PioC is a HiPIP (High Potential Iron Protein) and thus, it contains a 4Fe-4S cluster with a very high reduction potential ($E^0$= +450 mV vs SHE) being stable in the $[Fe_4S_4]^{2+}$ oxidation state. The protein has 54 amino acids and it is the smallest HiPIP ever isolated. The structure is unknown, but homology modelling with other HiPIPs suggests that the protein has a compact globular structure. It is characterized by the absence of topologically relevant secondary structure elements and is essentially composed by a series of loops and turns wrapped around the 4Fe-4S cluster[44,45]. The electronic properties of $[Fe_4S_4]^{2+}$ in HiPIPs have been studied in detail since about 40 years [4,46-52]. The magnetic coupling within the $Fe_4S_4$ cluster makes the electronic correlation times of the individual iron ions much shorter than isolated high spin $Fe^{3+}$ or $Fe^{2+}$ ion; nevertheless, paramagnetic contributions to nuclear relaxation are significant for nuclei within a 10 Å sphere from the cluster[53]. Therefore, PioC is a suitable system to address an important issue for inorganic structural biologists: can we improve methods for measuring relaxation rates in paramagnetic systems and use relaxation based NMR restraints [54] (or PREs, as they are now commonly defined [40,55]) as the sole source of dipole-dipole restraints? Can we break the paradigm that NOEs are an essential step for solution structure calculation?

A standard $^{15}$N HSQC experiment on a PioC sample shows only 39 amide resonances out of expected 49 non-Proline residues; however a $^{15}$N IR-HSQC-AP experiment shows additional 10 resonances, this demonstrated that all HN signals can be detected, provided that experiments specifically designed to observe fast relaxing resonances are used (**Figure 1**). In order to perform the complete resonance assignment of the protein, we combined the conventional approach based on triple resonance experiments (**Table S1**) with a non-systematic procedure using a combination of 1D NOEs, $^{13}$C direct detection, double and triple resonance experiments recorded with parameters optimized "a-la-carte" [56]. The non-systematic protocol for the complete NMR assignment is described in detail in **SI (Tables S2-S3** and **Figures S1-S4).** Eventually, these experiments provided the complete NMR assignment (**Table S4)**. We assigned (excluding the N-ter Val 1) 100% of backbone $^1H_N$, $^{13}$C, and $^{15}$N resonances, 98% of Hα, 86% and 91% of $^1$H and $^{13}$C side chains atoms, respectively. However, even if the $^1$H assignment is almost completed, $^{15}$N and $^{13}$C HSQC-NOESY experiments at high magnetic field gave only 344 meaningful NOEs that, without any additional information on cluster binding mode, were insufficient to obtain a structure. Three factors quench NOE intensities: i) the small rotational correlation time ($3.4 \times 10^{-9}$ s from $^{15}$N relaxation, **Figure S5**); ii) paramagnetic relaxation affecting at least 50% of the protein; iii) the absence of secondary structure elements, typical of HiPIPs. A few, crucial, structural constraints were obtained from residues binding the cluster, covalently or via H-bonds. Cys β$CH_2$ hyperfine shifts were converted into four $\chi_2$ dihedral angle constraints defining the cluster binding topology[57], seven crucial 1D NOEs between Cys β$CH_2$ and neighboring residues were quantified (**Figure S2**), large $^{15}$N shifts values,

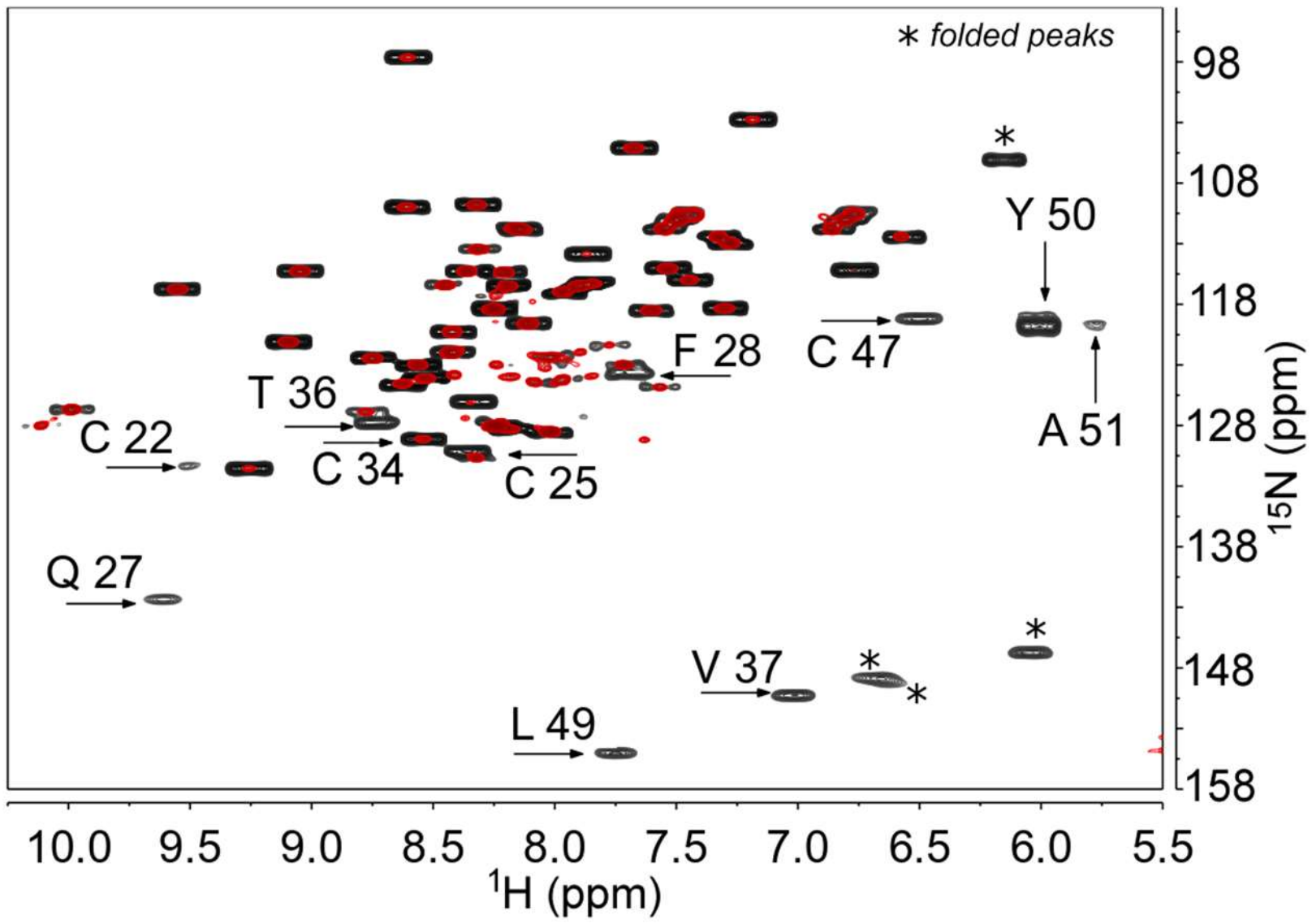

**Figure 1**. 500 MHz 298K, $^{15}$N- HSQC spectrum on PioC collected using the HSQC-AP experiment (black), overlaid with a standard $^{15}$N HSQC spectrum (red). Labeled signals are observable only in the HSQC-AP spectrum.

observed for Gln27, Val37 and Leu49 were taken as an evidence of three H-bonds, respectively linking HN to Sγ of the preceding (i-2 or i-3) cluster-bound Cys. These fourteen constraints were included into structure calculation, together with the geometrical parameters of the cluster, obtained as previously de-

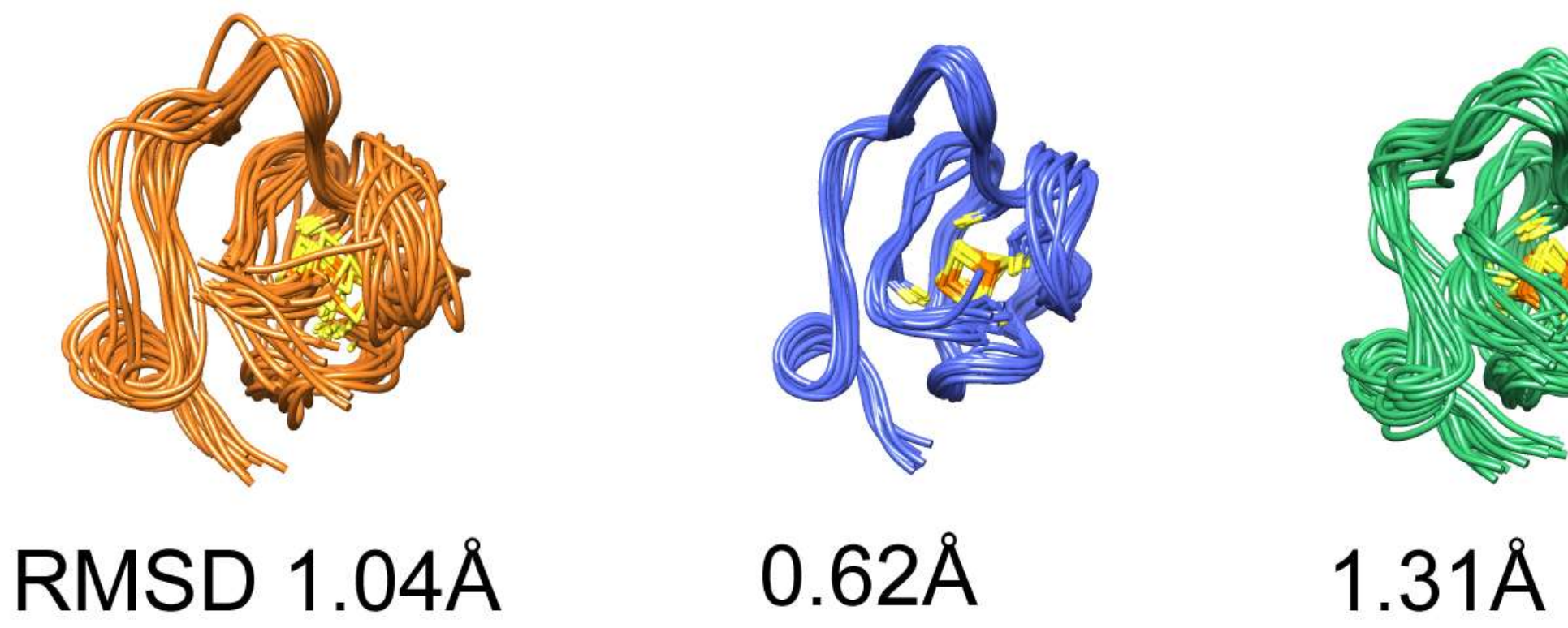

**Figure 2**. Solution structure of PioC obtained using NOEs only (orange), the full set of NMR restraints (blu), PREs only (green). In all cases, the families of 20 conformers were obtained from Torsion Angle Dynamics (CYANA2.1) and refinement using molecular dynamics (AMBER-16 package).

scribed [19]. We obtained, using torsion angle dynamics and refinement via molecular dynamics (see **SI** for details), a structure, shown in **Figure 2A,** with backbone and heavy atoms RMSD of 1.04 ±0.30 Å and 1.81±0.30 Å (residues 5-50). As expected, the cluster is the essential element to drive the fold of the polypeptide chain and the 14 constraints from cluster-bound residues are important to frame the cluster within the protein. Indeed, without them, we obtained structures with significantly higher backbone and heavy atoms RMSD of 1.27±0.20 Å and 1.95±0.20 Å. We now considered PREs: to this end, a novel $R_2$-weighted $^{15}$N-HSQC-AP experiments (**Figure S6**) was developed and used, together with the already described $^{15}$N-IR-HSQC AP experiment[8], to measure $R_1$ and $R_2$ values of all amide protons, as summarized in **Figure 3**. A $^{13}$C version of the IR-HSQC-AP experiment was also developed and allowed to record $^1$H relaxation rates for Hα backbone protons as well as for aliphatic and aromatic side chains

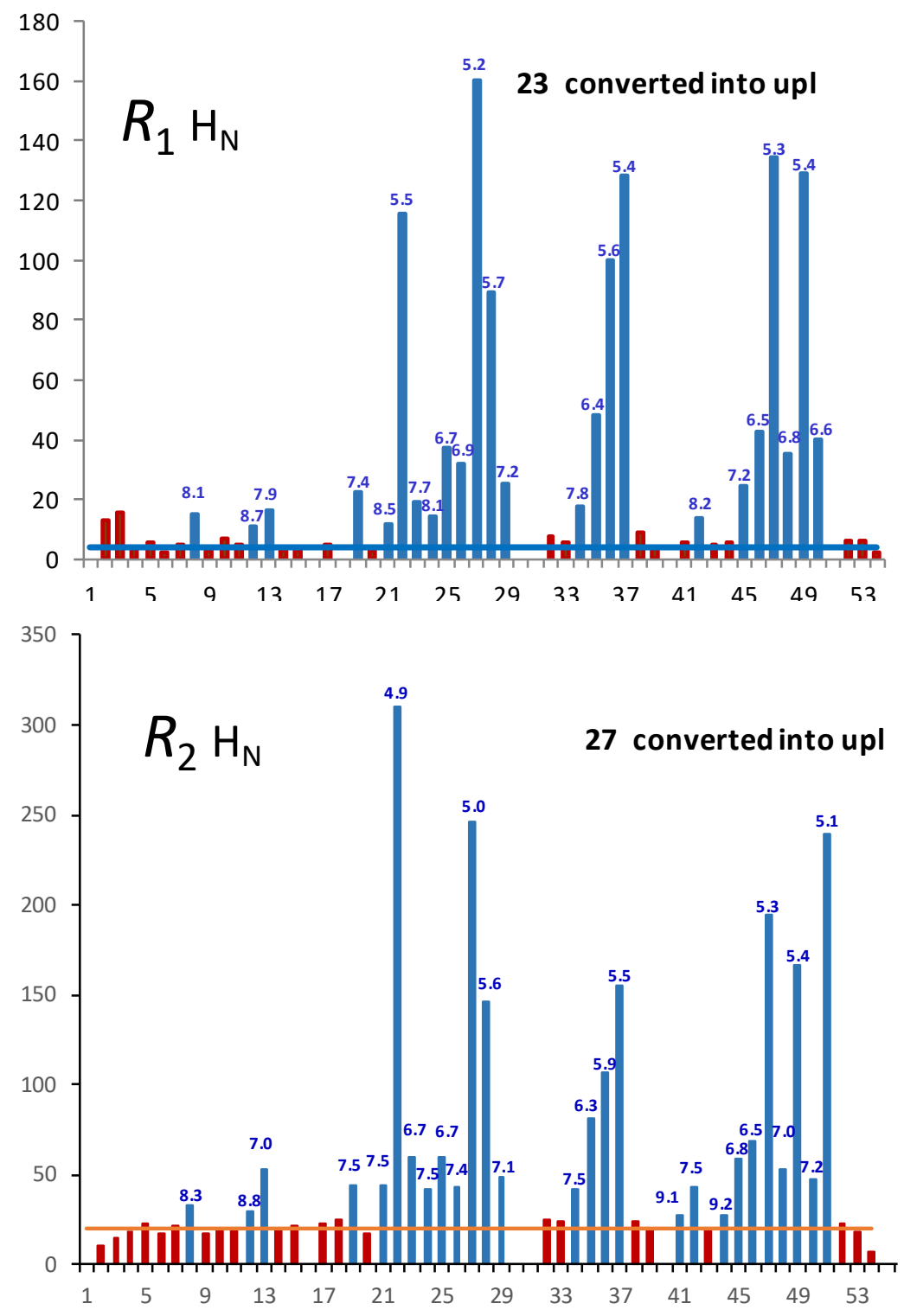

**Figure 3**. Longitudinal (upper) and transverse (lower) relaxation rates of amide $H_N$ signals. Horizontal lines show the threshold values taken as average diamagnetic value. Rates exceeding the threshold (blue-colored histograms) were converted into upper limit distance values, reported in Angstroem for each histogram.

(**Figure S7**). Finally, $^{1}H$ and $^{13}C$ resonances of cluster bound Cys residues were identified and assigned using rapid recycling experiments and from these twelve $R_1$ and $R_2$ values were obtained from inversion recovery and linewidth analysis of one dimensional $^{1}H$ and $^{13}C$ experiments. Overall, 312 relaxation rates, amounting to ca. six rates per residue, were measured, as summarized in **Table S7**. From the full set of relaxation data, $R_{1para}$ and $R_{2para}$ were found and converted into upper distance limits, following a procedure described in **SI**. In the case of $^{1}H_N$ signals, for which both $R_1$ and $R_2$ were available, the conversion of $R_{1\ para}$ and $R_{2para}$ into upper distance constraints was done, independently for each of the two measurements, according to an $r^{-6}$ dependence. Where different upper limit distances were found from $R_1$ and $R_2$, the less restrictive among them is considered, in order to minimize errors in upper limit conversion due to internal motions that may affect $R_1$ and $R_2$ to a different extent. For amide $H_N$ resonances, 33 upper distance limits were obtained out of 49 non-proline residues, indicating that about 66% of the protein is affected by paramagnetism, while 137 upper distance restraints are obtained from $H_C$ $R_1$ values and 5 from $^{13}C$ $R_1$ values. Overall, 175 constraints were obtained from relaxation rates. The number of distance constraints was increased to a total of 533. The use of relaxation based NMR restraints, together with those used previously (the summary of constraints is reported in **Table S5**), gave a high resolution NMR structure with backbone and side chain RMSD values of 0.62±0.11Å and 1.14±0.13Å respectively (**Figure 2B**). PREs improve the quality of the structure not only in the proximity of the cluster but throughout the entire protein. The combination of both type of restraints led to statistical parameters that are indicative of a highly precise structure of a well-folded protein of small/medium size (**Table S6**).

We now address whether an NMR structure obtained without NOEs is able to achieve good accuracy and precision. Indeed, **Figure 2C** shows the family of structures obtained without the 344 NOEs from $^{13}C$ and $^{15}N$-NOESY-HSQC experiments. The structure has backbone and heavy atoms RMSD of 1.31±0.27Å and 2.00±0.32Å, respectively. The overall structure precision is obviously lower than that obtained in previous calculations but still lies within an acceptable quality range. The per-residue comparison of backbone RMSD (**Figure 4A)** shows that the family obtained with the full set of constraints has always the lowest RMSD (but Thr24), indicating that the combination of NOEs and PREs improves the precision in all the protein regions. PREs provide information exactly were NOEs are missing, thus complementing NOE data. In several protein regions, the NOE-only family has an RSMD similar to the family obtained with the full set of constraints, indicating that NOEs drive the structure towards a minimum. Conversely, in other regions the precision is significantly improved by the use of PREs, thus we can define them as PRE driven. The loop surrounding the cluster and containing Cys 22 and Cys 25 of the CXXC binding motif in HiPIPs has a different trend. Here, the RMSD values of the three families are similar and higher than average values. This is the situation in which, not only NOEs but also PREs are missing due to the close proximity to the paramagnetic center. Essentially, for this fragment the structure is given by the cluster binding topology, by the dihedral angles of Cys bound residues and by H-bonded residues. We can obtain clues on the accuracy of the structures by comparing the three mean structures, the *PRE-only*, the

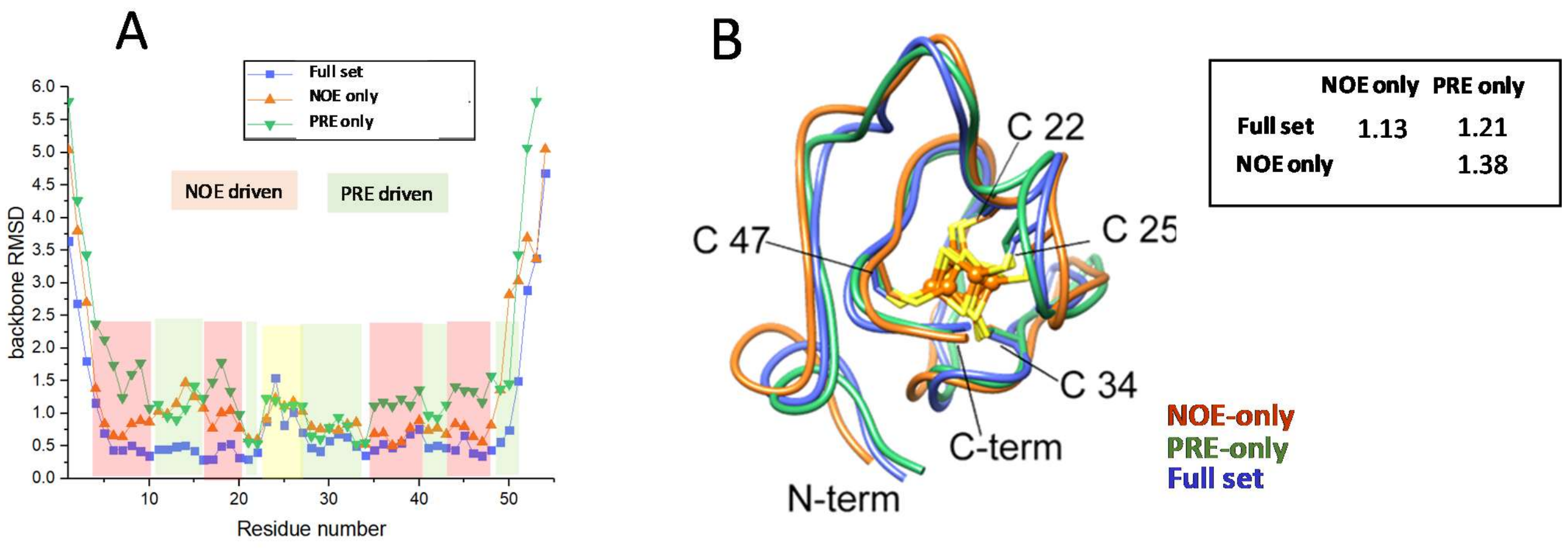

**Figure 4**. **A** per residue RMSD values of the three different families. The relative contribution of the different set of constraints on a per residue basis is shown with the color code indicated in the Figure. **B** Superimposing of the mean structures obtained with different sets of constraints. Figure reports also pairwise backbone RMSD values.

*NOE-only* and the *full-set* structures. As shown in **Figure 4B,** the pairwise RMSD between them are all similar and comparable with the precision within each family. The PRE-only structure is essentially the same, although with a larger RMSD, as the one obtained with the full-set of constraints. The latter is, for most of the protein, in an average position between PRE-only and NOE-only.

Notwithstanding the exciting perspectives opened by computational biologists[58-61], the quest for novel NMR restraints is still of primary importance in structural biology. Up to date, a dense network of NOEs has always been considered essential for structures, because constraints between residues that are far apart in the primary sequence define the relative orientation of different structural motifs[62]. Many authors introduced the use of PREs in this context and evoked the idea that they do have the property to replace NOEs in those protein regions where the paramagnetic effect is operative. However, until now, there are no cases reported in which NOE could be completely replaced by PREs. The NMR structure of PioC is a proof of concept of the above propositions: when a protein is small, enough to be affected by paramagnetism in a large percentage, then NOEs are not essential anymore, if relaxation rates are measured virtually for all $^1$H spins. In this case, an extended network where all the $^1$H spins are linked to a single point (the metal center) via long-range dipolar couplings can completely replace a network of short-range dipole-dipole $^1$H-$^1$H couplings. Noteworthy, PREs are unique also with respect to other paramagnetism-based restraints such as pseudocontact shifts and residual dipolar couplings. Noteworthy, PREs are unique also with respect to other paramagnetism-based restraints such as pseudocontact shifts and residual dipolar couplings[12]. The latter have a combination of angular and spatial dependence and thus, even when a large number of them are available, they are not able to replace NOEs completely. In PioC, the $[Fe_4S_4]^{2+}$ cluster provides upper distance limits up to 13 Å, while the average protein radius is about 15 Å, thus being an ideal case for obtaining the first NOE-less NMR structure. Besides the serendipity case of PioC, these results suggest the systematic use of PREs in structure calculations of metalloproteins because they provide distance restraints in protein regions where NOEs are sparse due to paramagnetism. Factors such as protein size, electronic correlation times of metal ion(s) and internal mobility will modulate the interplay between paramagnetism based and conventional NMR restraints and their relative contribution to the final structure. Whatever the combination will be, we may anticipate that, a combination of the largest possible set of constraints from both paramagnetic and conventional restraints will be the optimal solution in terms of structure quality.

## DEPOSITION

The 20 conformers with the lowest target function constituted the final family and the chemical shift assignment were deposited to the Protein Data Bank (PDBID 6XYV) and to the BMRB (code 34487) respectively.

## ACKNOWLEDGMENT

We are grateful to Prof. Gaetano Montelione for the enlightening discussions during his stay at CERM as invited Professor.

This work benefited from access to CERM/CIRMMP, the Instruct-ERIC Italy centre. Financial support was provided by European EC Horizon2020 TIMB3 (Project 810856) Instruct-ERIC (PID 4509).

This article is based upon work from COST Action CA15133, supported by COST (European Cooperation in Science and Technology). Fondazione Ente Cassa di Risparmio di Firenze (CRF 2016 0985) is acknowledged for providing fellowship to MI. Project LISBOA-01-0145-FEDER-007660 (Microbiologia Molecular, Estrutural e Celular) funded by FEDER funds through COMPETE2020 - Programa Operacional Competitividade e Internacionalização (POCI), Fundação para a Ciência e a Tecnologia (FCT) Portugal Grant PD/BD/135187/2017 to IBT

# NOE-less protein structures-an alternative paradigm in Highly Paramagnetic Systems

Inês Trindade[§], Michele Invernici[£], Francesca Cantini[£], Ricardo Louro[§]*, Mario Piccioli[£]*

[§] Instituto de Tecnologia Química e Biológica António Xavier (ITQB-NOVA), Universidade Nova de Lisboa, Av. da República (EAN), 2780-157 Oeiras, Portugal, louro@itqb.unl.pt.

[£] Magnetic Resonance Center and Department of Chemistry, University of Florence, Via L. Sacconi 6 50019 Sesto Fiorentino, Italy.

# Supplementary Material

### 1.Protein expression and purification.

PioC was expressed and purified as previously reported [1]. Three samples of PioC were produced (unlabeled, single $^{15}$N-labeled, double $^{13}$C &$^{15}$N-labeled) and the expression and purification protocol was identical throughout except in the addition of ammonium sulfate ($^{15}N_2$, 99%) and [U-$^{13}C_6$] D-glucose in the M9 minimal media when labeling was required. BL21 DE3 cells were double transformed with pET32h, a plasmid containing the construct thioredoxin–6xHis–thrombin cleavage site–PioC, and with pDB1281, a plasmid that carries the machinery for the assembly of iron-sulfur clusters. Cells were grown in Luria-Bertani (LB) supplemented with 100mg*dm$^{-3}$ ampicillin and 35mg*dm$^{-3}$ chloramphenicol until the $OD_{600nm}$ of 0.6 where they were induced with arabinose and $FeCl_3$ and cysteine were added. Cells were again incubated until the $OD_{600nm}$ of 1 and then harvested and washed in M9 minimal media salts before resuspension in M9 minimal media. Once re-suspended, cells were incubated for one hour before induction with 0.5mM IPTG. After 4hr cells were harvested by centrifugation and disrupted using a French Press at 1000psi. The lysate was ultra-centrifuged at 204 709g for 90 min at 4ºC to remove cell membranes and debris and the supernatant was dialyzed overnight against 50 mM potassium phosphate buffer pH 8 with 300mM NaCl before injection in a His-trap affinity column (GE Healthcare). The fraction containing Histag-PioC eluted with 250mM imidazole and was incubated overnight with Thrombin (GE Healthcare) for digestion. The final purified PioC (His-tag free) was then concentrated from the flow through of a 2nd passage through the His-trap column using an Amicon Ultra Centrifugal Filter (Millipore) with a 3kDa cutoff. The purity of PioC was confirmed by SDS-PAGE with Blue Safe staining (NzyTech) and by UV-Visible spectroscopy.

## 2.NMR experiments

All experiments were recorded using Bruker AVANCE-NEO spectrometers, equipped with cryogenically cooled triple resonance inverse detection probeheads (CP-TXI), except $^{13}C$-detected experiments, which were acquired at 176.05 MHz using a cryogenically cooled probehead optimized for $^{13}C$ direct detection (CP-TXO), and $^{1}H$ experiments which were recorded at 400 MHz using a room temperature, selective 5mm $^{1}H$ probe without pulsed field gradients. All spectra were processed using the Bruker software TopSpin. Standard radio frequency pulses and carrier frequencies for triple resonance experiments were used. The set of NMR experiments used for sequence specific assignment, NOE collection and $^{15}N$ relaxation analysis is summarized in **Table S1**.

**Table S1**

| Experiments | Time Domain Data Size (points) | | | Spectral width (ppm) | | | ns | Delay time (s) | Magnetic field (MHz) |
|---|---|---|---|---|---|---|---|---|---|
| | $t_1$ | $t_2$ | $t_3$ | $F_1$ | $F_2$ | $F_3$ | | | |
| **[$^{1}H$-$^{1}H$]-NOESY[a]** | 1024 | 2048 | | 14.6 ($^{1}H$) | 14.6 ($^{1}H$) | | 48 | 1.2 | 900 |
| **[$^{1}H$-$^{1}H$]-TOCSY** | 600 | 2048 | | 14.0 ($^{1}H$) | 14.0 ($^{1}H$) | | 48 | 2.0 | 600 |
| **$^{1}H$-$^{15}N$-HSQC** | 128 | 2048 | | 40.0 ($^{15}N$) | 13.2 ($^{1}H$) | | 2 | 1.5 | 500 |
| **$^{1}H$-$^{13}C$-HSQC** | 320 | 1024 | | 80.0 ($^{13}C$) | 13.0 ($^{1}H$) | | 20 | 2.0 | 600 |
| **cbcaconh** | 128 | 48 | 1800 | 80.0 ($^{13}C$) | 40.0 ($^{15}N$) | 13.2 ($^{1}H$) | 8 | 1.3 | 900 |
| **cbcanh** | 128 | 48 | 1800 | 80.0 ($^{13}C$) | 40.0 ($^{15}N$) | 13.2 ($^{1}H$) | 16 | 1.3 | 900 |
| **hnco** | 72 | 48 | 1800 | 16.0 ($^{13}C$) | 40.0 ($^{15}N$) | 13.2 ($^{1}H$) | 8 | 1.3 | 900 |
| **hncaco** | 72 | 48 | 1800 | 16.0 ($^{13}C$) | 40.0 ($^{15}N$) | 13.2 ($^{1}H$) | 24 | 1.3 | 900 |
| **hnca** | 96 | 48 | 1800 | 40.0 ($^{13}C$) | 40.0 ($^{15}N$) | 13.2 ($^{1}H$) | 16 | 1.3 | 900 |
| **hbhanh** | 128 | 40 | 2048 | 13.2 ($^{1}H$) | 40.0 ($^{15}N$) | 13.2 ($^{1}H$) | 16 | 1.3 | 900 |
| **hnha** | 48 | 128 | 2048 | 40.0 ($^{15}N$) | 13.2 ($^{1}H$) | 13.2 ($^{1}H$) | 16 | 1.3 | 900 |
| **(H)CCH-TOCSY** | 128 | 64 | 1800 | 80.0 ($^{13}C$) | 80.0 ($^{15}N$) | 13.2 ($^{1}H$) | 16 | 1.5 | 900 |
| **$^{15}N$-edited [$^{1}H$-$^{1}H$]-NOESY** | 128 | 40 | 2048 | 13.2 ($^{1}H$) | 40.0 ($^{15}N$) | 13.2 ($^{1}H$) | 32 | 1.2 | 900 |
| **$^{13}C$-edited [$^{1}H$-$^{1}H$]-NOESY** | 192 | 64 | 2048 | 14.0 ($^{1}H$) | 80.0 ($^{13}C$) | 14.0 ($^{1}H$) | 32 | 1.2 | 950 |
| **$^{15}N$ $R_1$[b]** | 128 | 1024 | | 41.0 ($^{15}N$) | 15.6 ($^{1}H$) | | 16 | 3.0 | 500 |
| **$^{15}N$ $R_2$[c]** | 128 | 1024 | | 41.0 ($^{15}N$) | 15.6 ($^{1}H$) | | 16 | 3.0 | 500 |
| **$^{15}N$- NOE** | 144 | 1024 | | 41 | 15.6 | | 56 | 5.0 | 500 |
| **$^{1}HN$ $R_1$** | 220 | 1024 | | 80.0 ($^{15}N$) | 21.7 ($^{1}H$) | | 16 | 4.0 | 500 |
| **$^{1}HN$ $R_2$** | 156 | 1024 | | 80.0 ($^{15}N$) | 21.7 ($^{1}H$) | | 64 | 4.0 | 500 |
| **$^{1}HC$ $R_1$** | 320 | 1024 | | 80.0 ($^{13}C$) | 13.0 ($^{1}H$) | | 20 | 2.0 | 600 |
| **CON** | 128 | 1024 | | 40.6 ($^{15}N$) | 47.0 ($^{13}C$) | | 64 | 1.0 | 700 |

## 3.Paramagnetism-tailored NMR Experiments

To identify signals affected by the hyperfine interaction, tailored experiments were performed. Most of them are experiments, commonly used in biomolecular NMR spectroscopy, where the standard set of parameters is modified to account for the shift and relaxation properties of paramagnetic systems. The pulse sequences are modified to choose those building blocks that are more "robust" with respect to fast nuclear relaxation, and the magnetic field is chosen accordingly. A few experiments have been specifically designed to identify resonances/dipolar or scalar couplings/relaxation properties of paramagnetic systems [2-4]. For each of them, typical parameter choices are summarized in **Table S2**

**Table S2**

| Experi-<br>ments | Time Domain Data Size | | Spectral width | | ns | Delays | Magnetic field |
|---|---|---|---|---|---|---|---|
| | (points) | | (ppm) | | | (s) | (MHz) |
| | $t_1$ | $t_2$ | $F_1$ | $F_2$ | | | |
| **$^1$H-NOE** | 8192 | | 80.0<br>($^1$H) | | 400000 | aq= 133 ms<br>saturation=110ms | 400 |
| **IR- $^{15}$N-HSQC-AP** | 220 | 1024 | 80.0 | 21.7<br>($^1$H) | 256 | aq=47ms<br>recyc=0.15s<br>invrec=0.2s<br>$\delta$=700us | 500 |
| **$R_2$-weighted $^{15}$N HSQC-AP** | 256 | 1024 | 80.0<br>($^{15}$N) | 21.7<br>($^1$H) | 256 | aq=47ms<br>recyc=0.15s<br>$\delta$=2.0ms | 500 |
| **$^1$H-$^{13}$C-HNCA** | 160 | 1024 | 100.0<br>($^{13}$C) | 25.0<br>($^1$H) | 512 | aq=40ms<br>Recyc=0.5s<br>$\delta$=2.0ms | 500 |
| **$^1$H-$^{1\%}$N-HNCA** | 128 | 1024 | 70.0<br>($^{15}$N) | 25.0<br>($^1$H) | 512 | aq=40ms<br>Recyc=0.5s<br>$\delta$=2.0ms | 500 |
| **1d-$^{13}$C** | 8192 | | 245.6<br>($^{13}$C) | | 32768 | aq= 71ms<br>recyc=73ms | 900 |
| **[$^{13}$C-$^{13}$C]-COSY** | 500 (400) | 2048<br>(1500) | 258.2<br>($^{13}$C) | 258.2<br>($^{13}$C) | 480 | aq=22ms<br>recyc= 0.2s | 700 |
| **IR- $^{13}$C-HSQC-AP** | 256 | 2048 (512) | 136.0<br>($^{13}$C) | 60.2<br>($^1$H) | 1024 | aq=34ms<br>recyc=0.15s<br>$\delta$=2.0ms<br>Invrec=30ms | 500 |
| **$^1$HC $R_1$ IP acq** | 320 | 1024 | 136.0<br>($^{13}$C) | 60.2<br>($^1$H) | 320 | aq=17ms<br>recyc=0.12s<br>invrec=80ms<br>$\delta$=600us | 500 |
| **$^1$HC $R_1$ AP acq** | 320 | 1024 | 136.0<br>($^{13}$C) | 60.2<br>($^1$H) | 1024 | aq=17ms<br>recyc=34ms<br>invrec=80ms<br>$\delta$=600us | 500 |

## 4.Assignment of hyperfine shifted signals

In order to identify signals affected by the hyperfine interaction, we recorded a paramagnetic-tailored HNCA experiment (see previous section). Thanks to this HNCA experiment (**Figure S1A**), $H_N$, $N_H$ and Cα resonances of Cys22, Cys25 and Cys34 have been assigned due to the sequential connectivities observed with HN signals of residues Gly23, Arg26 and Ile35, unambiguously identified with the standard assignment strategy. **Figure S1B** shows the superimposition of two paramagnetic $^{13}$C-HSQC-AP experiments, at 288 K (black lines) and at 298 K (red lines). The Anti-Curie temperature dependence provides the identification of 4-β$CH_2$ and 3-αCH of the ligand cysteines. Through the cross peaks obtained by $^{13}$C-HSQC experiment and the cysteines Cα assigned by HNCA, the Hα resonances of Cys22, Cys25 and Cys34 have been assigned.

The $^{1}$H NMR spectrum (**Figure S2A**) contain two signals experiencing paramagnetic line broadening, downfield hyperfine shift (16.6 ppm and 14.5 ppm) and antiCurie temperature dependence, showing the typical fingerprint of FeS protein binding a $[Fe_4S_4]^{2+}$cluster, where the two peaks arise from Cys Hβ signals [5]. Traces B-C show the NOE difference experiments performed upon selective irradiation of signal at 16.6 ppm and 14.5 ppm, respectively. They allow us to detect and measure Nuclear Overhauser enhancements (NOEs) between these two protons and their environment. From the selective saturation of the Hβ signal at 16.6 ppm (**Figure S2B**), a strong NOE is observed with its geminal partner at 6.60 ppm (f). Other NOEs are observed with Gly 23 $H_N$ (c ), with Arg 21 Hβ (h) and Arg 21 Hγ (i); this network of inter-residue connectivities allows us to assign the signal at 16.6 ppm as the Cys 22 Hβ2 signal. The assignment of Cys 22 Hβ1 at 6.60 ppm is also confirmed by a cross peak, observed in a 2D NOESY (data not shown), between the peak at 6.60 ppm and the previously assigned Arg 21 Hε. Additionally, this NOE experiment also allows us to identify spatial proximity with Hδ and Hε of Phe 28 (e,f) and with Glu 43 Hα (g). Upon irradiation of Hβ signal at 14.5 ppm **(Figure S2C**), a strong NOE is observed to the HN resonance of Cys 34 at 8.55 ppm (j), already identified from the HNCA experiment and identifies signal b as the Hβ2 of Cys 34. In addition, Cys 34 Hβ2 (14.5 ppm) gives NOEs to Hα Val 37 and Hγ Ile 41. The assignment of Cys 22 and Cys 34 Hβ signals allows to identify the Cβ from the $^{13}$C HSQC-AP experiments shown in **Figure S1B**. HNCO experiments (data not shown) provide the assignment of all C' from the cross peaks with the related already assigned HN(i+1) nuclei.

$^{13}$C direct detection provides precious spectroscopic information. The 1D $^{13}$C experiments are shown in **Figure S3A**. Signals belonging to Cys bound residues are expected to experience a strong antiCurie Temperature dependence, this allows to confirm the assignment of Cys 22 and Cys 34 Cα and Cβ

signals and of Cys 25 Cα. Additionally, two other Cys Cβ signals are observed at 111.1 and 102.65 ppm. C-C COSY has proved to be a powerful technique able to identify all Cα/C' connectivities of cluster bond Cysteine. A tailored choice for $t_{1max}$ and $t_{2max}$ values provided [6], for the upper diagonal part, the spectrum shown in **Figure S3B,** which confirmed the assignment of Cα of cysteine 22, 25 and 34 and identified the missing Cα Cys 47, unobserved in the 1D $^{13}$C experiments because in overlap with Cys34 and Cys22 Cα and unobserved also in the $^{13}$C-HSQC-AP experiment shown previously. The set of $^{13}$C direct detected experiments is completed by a CON experiment (**Figure S3C**), which confirms the cysteine amide nitrogen assignment for Cys 22, 25 and 34 and provides the assignment for the amide nitrogen of Cys 47.

The information obtained from the paramagnetic $^{15}$N-HSQC-AP experiment have been already discussed in the text. In particular, the amide HN of Cys 47 escaped detection also in the "optimized" version of HNCA and HNCO experiments. This accounts for a very efficient $R_2$ $^1$H relaxation of the amide proton that would prevent the identification of its HNCA and HNCO peaks. Indeed, the $^{15}$N-HSQC-AP experiment shows a very broad, upfield shifted HN amide signal (6.47/118.9 ppm), with a $^{15}$N shift consistent with the assignment of Cys 47 amide, obtained with CON (C' of Trp 46 was assigned via HNCACO from the HN of Trp 46). Consistently, the $H_N$ signal at 6.50 ppm has $^1$H $T_2$ values of 5.1 ms, accounting for an HN belonging to the first coordination sphere of the cluster without any connectivity in triple resonance experiments and, therefore, it can be safely assigned as Cys 47. The four Cysteines HN peaks are pointed out in **Figure S4**. All of them are affected by paramagnetic relaxation; however paramagnetism affects these four signals at different extent, as qualitatively observed here from relative peak intensities in the $^{15}$N-HSQC-AP and quantitatively confirmed by $^1$H $R_1$ and $R_2$ relaxation data (see later). Cys 25 and Cys 34 have relatively sharp HN resonances, observable even in a HNCA experiment, suggesting a Cysteine orientation in which the amide group is pointing far from the cluster. On the other hand, HN of Cys 47 and of Cys 22 are observable in a $^{15}$N HSQC only at very short INEPT transfer delays, indicating that the HN vector is, in these cases, pointing towards the cluster. This is fully consistent with what observed in our NMR structure. Indeed, this feature is highly conserved throughout the very many HiPIPs of known structure [7-11].

The two $\beta CH_2$ protons of Cys 25 and Cys 47 can be discriminated by analyzing relaxation rates of Hβ protons, as measured from a $^{13}$C IR-HSQCP-AP experiment. The $^{13}$C HSQC-AP shows the $\beta CH_2$ groups (**Figure S2B**) of Cys 25 and Cys 47 that are not yet sequence specifically assigned. Measured $R_1$ $^1$H rates are 313 $s^{-1}$ and 282 $s^{-1}$ for the two protons bound to the Cβ at 103.8 ppm and 450 $s^{-1}$ and 130 $s^{-1}$ for the two protons bound to the Cβ at 88.3 ppm. Conversion of $R_1$ $^1$H rates into upper distance limits gives values of 3.8 Å and 3.9 Å for the two protons bound to the Cβ at 103.8 pm, while the two protons bound to the Cβ at 88.3 ppm give metal-to-proton distances of 3.6 Å and 4.5 Å. The inspection of X-ray and NMR structures of the many HiPIPs characterized so far, invariantly shows that the cluster binding topology of Cys II give rise to very similar metal-to-proton distances for the two Hβ protons, while the two Hβ of Cys IV are always asymmetric, with one Hβ close to the cluster and the other much farther apart. This is a strong indication that the symmetrical $\beta CH_2$ group with Cβ at 103.8 ppm belongs to Cys 25 while $\beta CH_2$ protons with asymmetric distances, with Cβ at 88.3 ppm, are assigned to Cys 47. Overall, the assignment of Cys resonances and the key experiments supporting their assignment is summarized in **Table S3.**

**Table S3.** Sequence-Specific Assignments of Cysteine resonances. For each resonance, color code indicates the experiment used to perform the assignment.

| | Chemical shift (ppm) | | | | | | | |
|---|---|---|---|---|---|---|---|---|
| | **HN** | **N** | **Cα** | **Hα** | **Cβ** | **Hβ1** | **Hβ2** | **CO** |
| **Cys22** | 9.47 | 131.4 | 89.5 | 8.18 | 102.65 | 6.6 | 16.6 | 177.6 |
| **Cys25** | 8.36 | 130.3 | 80.0 | 3.82 | 103.8 | 8.09 | 9.94 | 176.5 |
| **Cys34** | 8.55 | 129.1 | 90.3 | 3.83 | 111.1 | 5.8 | 14.5 | 173.9 |
| **Cys47** | 6.50 | 118.8 | 90.3 | | 88.3 | 7.99 | 8.95 | 170.7 |

*Table legend*

| | | | | |
|---|---|---|---|---|
| HNCA | $^{13}$CHSQC | 1D-NOE | 2D-NOESY | Relaxation analysis |
| C-C COSY | CON | $^{15}$N-HSQC | HNCO | |

## 4.Sequence Specific Assignment

Data analysis and resonances assignment were performed using CYANA 2.1. Proton resonances were calibrated with respect to the signal of 2,2-dimethylsilapentane-5-sulfonic acid (DSS). Nitrogen chemical shifts were referenced indirectly to the $^{1}H$ standard using a conversion factor derived from the ratio of NMR frequencies. Carbon resonances were calibrated using the signal of dioxane at 69.4 ppm (298 K) as secondary reference. The complete assignment is reported in **Table S4**. All parameters of the experiments used are reported in Tables S1 and S2.

**Table S4** $^{1}H$, $^{15}N$ and $^{13}C$ resonance assignments for PioC at 298 K, in 50 mM phosphate, 300 mM KCl at pH 5.8.

| aa | N (HN) | CO | Cα (Hα) | Cβ (Hβ) | Others |
|---|---|---|---|---|---|
| Val-1 | | 176.34 | 62.36 | 32.34 | |
| Thr-2 | 117.2 (8.26) | 174.0 | 61.8 (4.26) | 69.78 (4.12) | 21.67 (1.13) |
| Lys-3 | 123.7 (8.42) | 176.0 | 55.7 (4.32) | 33.0 (1.69) | Cγ 24.3 (1.38) Cδ 28.9 (1.63) Cε 41.68 (2.96) |
| Lys-4 | 121.8 (8.43) | 176.0 | 56.7 (4.14) | 34.7 (1.45, 1.37) | Cγ 26.4 (1.32,0.95) Cδ 29.9 (-) Cε 42.3 (2.91, 2.74) |
| Ala-5 | 122.3 (8.76) | 177.5 | 51.4 (4.51) | 19.2 (1.56) | |
| Ser-6 | 115.2 (9.08) | 175.0 | 56.9 (4.41) | 64.9 (4.32, 4.07) | |
| His-7 | 116.3 (9.64) | 177.6 | 58.7 (4.24) | 28.2 (3.39, 3.32) | Cδ2 117.5 (6.69) Cε1 136.8 (8.24) |
| Lys-8 | 116.5 (8.49) | 179.6 | 59.4 (4.19) | 32.4 (1.90, 1.74) | Cγ 24.3 (1.49,1.41) Cδ 29.2 (1.71, 1.69) Cε 41.7 (2.98) |
| Asp-9 | 119.5 (8.13) | 178.2 | 57.2 (4.36) | 40.0 (2.85, 2.71) | |
| Ala-10 | 118.2 (8.23) | 176.2 | 55.7 (4.28) | 27.2 (1.91) | |
| Gly-11 | 105.0 (7.69) | 175.1 | 46.8 (3.90, 3.73) | | |
| Tyr-12 | 116.3 (8.19) | 174.8 | 59.1 (4.51) | 39.3 (2.56, 2.37) | Cδ 135.25 (6.88) Cε 120,59 (7.33) |
| Gln-13 | 124.5 (8.67) | 172.93 | 52.6 (4.66) | 32.5 (2.11, 1.92) | Cγ 32.7 (2.48, 2.77) Nε2 112.36 (7.35, 6.58) |
| Glu-14 | 115.2 (8.22) | 174.7 | 56.3 (4.09) | 29.1 (2.28, 1.81) | Cγ 36.8 (2.35) |
| Ser-15 | 112.8 (7.29) | 171.8 | 56.2 (4.99) | 63.9 (3.73. 3.68) | |
| Pro-16 | 130.36 | 176.1 | 63.0 (4.15) | 33.3 (1.94,1.68) | Cγ 27.6 (1.87,1.69) Cδ 51.3 (3.40, 4.03) |
| Asn-17 | 116.4 (7.92) | 175.7 | 50.0 (4.41) | 36.2 (1.04, 0.69) | Nδ2 106.15 (6.17, 3.93) |
| Gly-18 | 111.6 (8.17) | 174.1 | 46.6 (3.38,3.71) | | |
| Ala-19 | 126.9 (8.78) | 178.0 | 52.2 (4.21) | 19.2 (1.3) | |
| Lys-20 | 118.4 (7.62) | 173.1 | 56.5 (3.86) | 32.1 (1.54, 0.26) | Cγ 24.5 (1.30, 1.13) Cδ 29.2 (1.59, 1.40) Cε 42.0 (2.95, 2.94) |
| Arg-21 | 115.87 (7.47) | 176.8 | 53.5 (6.13) | 32.9 (2.43, 1.80) | Cγ 26.5 (1.69, 1.45) Cδ 43.6 (3.50, 3.00) Nε 84.94 (7.34) |
| Cys-22 | 131.4 (9.47) | 177.6 | 89.5 (8.18) | 102.65 (6.60, 16.6) | |
| Gly-23 | 97.54 (8.62) | 176.0 | 46.2 (3.91,3.81) | | |
| Thr-24 | 109.8 (8.62) | 173.6 | 60.8 (4.69) | 69.1 (4.95) | Cγ2 21.03 (1.18) |
| Cys-25 | 130.0 (8.36) | 176.5 | 80.0 (3.86) | 103.8 (9.94,8.09)* | |
| Arg-26 | 131.4 (9.28) | 179.3 | 58.9 (4.46) | 31.0 (1.94,1.84) | Cγ 27.1 (1.72,1.65) Cδ 43.5 (3.15, 3.09) Nε 82.5 (6.96) |
| Gln-27 | 141.7 (9.62) | 175.4 | 59.5 (4.36) | 32.0 (2.36) | Cγ 28.0 (2.47, 1.99) |
| Phe-28 | 123.4 (7.72) | 178.0 | 63.7 (4.40) | 52.3 (3.21) | Cδ 134.0 (7.13) Cε 130.38 (7.0) Cζ 129.16 (7.12) |
| Arg-29 | 128.2 (8.19) | 171.1 | 49.9 (4.59) | 29.9 (1.28, 1.13) | Cγ 25.5 (1.49, 1.40) Cδ 42.8 (3.34, 2.91) Nε 83.3 (7.22) |
| Pro-30 | | 176.2 | 61.2 (3.39) | 30.3 (2.23, 1.71) | Cγ 26.2 (1.88, 1.78) Cδ 50.5 (3.52, 2.79) |
| Pro-31 | 128.9 | 177.9 | 64.3 (4.40) | 35.0 (2.40, 2.12) | Cγ 24.5 (1.91, 1.85) Cδ 50.3 (3.59, 3.33) |
| Ser-32 | 111.8 (8.15) | 174.34 | 57.4 (5.43) | 64.5 (4.05, 3.83) | |
| Ser-33 | 114.9 (7.55) | 169.5 | 57.2 (4.69) | 66.0 (3.57, 3.47) | |
| Cys-34 | 129.0 (8.55) | 173.8 | 90.0 (3.83) | 111.1 (14.48,5.75) | |
| Ile-35 | 126.0 (8.36) | 178.0 | 64.1 (4.16) | 38.5 (2.03) | Cδ1 12.9 (0.91) Cγ1 26.9 (1.67, 1.46) Cγ2 18.8 (1.00) |
| Thr-36 | 127.6 (8.75) | 173.1 | 62.3 (4.30) | 69.7 (4.06) | Cγ2 21.7 (1.16) |
| Val-37 | 149.9 (7.00) | 173.9 | 63.0 (5.02) | 36.8 | Cγ1 26.0 (--) Cγ2 23.8 (1.24) |
| Glu-38 | 122.8 (8.61) | 174.8 | 56.9 (3.79) | 31.1 (1.82, 1.73) | Cγ 36.2 (2.19, 2.15) |
| Ser-39 | 115.7 (8.43) | 171.4 | 59.0 (4.06) | 62.7(4.07, 3.54) | |
| Pro-40 | 136.1 | 174.3 | 62.6 (4.58) | 34.8 (2.34, 2.02) | Cγ 24.2 (1.92, 1.89) Cδ 50.6 (3.41, 3.28) |
| Ile-41 | 118.2 (7.28) | 175.5 | 58.2 (4.10) | 39.3 (1.34) | Cδ1 20.8 (0.89) Cγ1 26.9 (1.04, 0.28) Cγ2 21.0 (0.09) |
| Ser-42 | 116.4 (7.84) | 178.3 | 55.2 (4.87) | 65.0 (3.96,3.64) | |
| Glu-43 | 121.2 (9.12) | 176.0 | 58.7 (3.32) | 29.2 (2.11, 1.96) | Cγ 34.9 (2.24, 2.04) |
| Asn-44 | 116.8 (7.98) | 172.2 | 52.9 (5.00) | 40.1 (3.00, 2.42) | Nδ2 111.7 (7.56, 6.88) |
| Gly-45 | 102.6 (7.20) | 173.0 | 45.5(4.14,3.77) | | |
| Trp-46 | 113.7 (7.86) | 172.6 | 65.5 (5.19) | 31.1 (2.85, 3.87) | Nε1 124.0 (8.53) Cδ1 126.0 (6.66) Cζ2 114.72 (7.39) Cζ3 122.07 (6.91) Cη2 124.26 (6.88) Cε3 119.9 (6.74) Cε2 139.48 (--) |
| Cys-47 | 118.9 (6.47) | 170.6 | 90.0 | 88.3 (7.99,8.95) | |
| Arg-48 | 115.2 (6.79) | 180.22 | 57.4 (3.60) | 29.8(1.62) | Cγ 27.9 (1.53, 1.66) Cδ 43.5 (3.18, 3.11) Nε 82.4 (7.26) |
| Leu-49 | 154.7 (7.74) | 180.9 | 58.2 (4.13) | 42.4 | Cγ 30.2 (1.30) 34.8 (1.40) 21.9 |
| Tyr-50 | 119.5 (6.0) | 173.7 | 62.4 | | Cδ 137.2 (6.30) Cε 120.1 (7.15) |
| Ala-51 | 119.0 (5.46) | 176.09 | 50.8 (4.26) | 20.7 (1.01) | |
| Gly-52 | 109.7 (8.33) | 174.5 | 45.8 (3.68,3.99) | | |
| Lys-53 | 120.2 (8.43) | 175.54 | 57.0 (4.14) | 33.4 (1.79) | Cγ 25.2 (1.38) Cδ 29.7 (1.66) Cε 38.6 (-) |
| Ala-54 | 128.5 (8.04) | 182.5 | 53.8(4.10) | 20.2(1.29) | |
| *Stereo specific assignment of $\beta CH_2$ not available | | | | | |

## 5. Structural constraints and structure calculation

Diamagnetic NMR restraints. NOEs were analyzed and converted into upper distance limits and used for manual structure calculation in CYANA 2.1. Backbone dihedral angle constraints were derived from $^{15}$N, $^{13}$C', $^{13}$C$\alpha$,$^{13}$C$\beta$, and H$\alpha$ chemical shifts, using TALOS+ and added as restraints. Overall, 344 meaningful upper distance constraints and 51 dihedral angles were used to calculate the structure. The data are summarized in **Table S5**.

Paramagnetic NMR restraints. Hyperfine shifts of Cysteines $\beta CH_2$ protons were converted into Cysteines dihedral angles $\chi_2$ according to the procedure already described [12,13]. 1D NOEs observed from well resolved, hyperfine shifted signals were measured according to a procedure originally describe in [14] and converted into upper distance limits provided the known distance among geminal $\beta CH_2$ protons.

Hydrogen bond donor atoms were identified by considering $^{15}$N shift values that are outliers by more than 20 ppm the average values according to BMRB Data Bank. Three H-bonding donor HN groups were identified. The three acceptors sulfur atoms were unambiguously identified from the NMR structure obtained without H-bonds and the three H-bonds added as structural constraints.

$R_1$ and $R_2$ rates of $^1H_N$ and $^1H_C$ were converted into upper distance limits as described in the following section. When $R_1$ and $R_2$ provided different upper distance limits for the same $^1H_N$ proton, the upper limit value was taken by considering the less restrictive value among the two. In these cases, the upper limit value was given a weighting factor 2. Overall, 39 values were taken from $H_N$, out of which 19 were weighted by a factor 2; 31 upper distance limits were taken from H$\alpha$ and 90 upper distance limits were taken from side chains (HC groups). Additionally, 10 upper limit values were taken by relaxation rate measurements of $\beta CH_2$ and $\alpha$CH from the four cluster-bound Cysteines. The total number of $^1$H based PRE restraints was 170. In addition, the relaxation rates of Cysteines C$\alpha$ and C$\beta$ carbon atoms were also measured and converted into 5 additional PRE based restraints. Overall, the number of relaxation based restraints used into structure calculation was 175, as summarized in **Table S5**.

Definition of the cluster for structure calculations. The Iron sulfur cluster was inserted into structure calculation according to the procedure originally described [10]. A special residue, named CFS, was added to the CYANA library. The artificial residue, denoted CFS, consists of a cysteinyl residue in which the thiol hydrogen (H$\gamma$) was replaced by an iron atom (Fe$\delta$) at the proper distance and by adding to the latter, through another covalent bond, the sulfur atom (S$\varepsilon$) constituting the inorganic sulfide of the cluster. Bond lengths and angles used in this construction were taken from previously reported structures [10,15,16]. Eight additional covalent bonds were added as link statements to the end of the sequence file between each iron atoms (Fe) and the two bonded sulfur atoms (*S*ε). This removes the van der Waals interactions between the Fe and the other ligands. Then, upper and lower distance limits are imposed along the eight edges of the cubane (the remaining four are defined within the four CFS residues), along the six edges of the tetrahedron described by the four iron atoms, six others along the edges of the tetrahedron described by the four inorganic sulfur atoms (*S*ε) and finally, among the six edges of the tetrahedron formed by cysteine *S*γ atoms. A total of 26 upper and lower distances limits was used in the CYANA calculations. This construction allows us to define a rigid cluster while leaving undefined the chirality of the peptide folding around it. The summary of conformationally restricting constraints is reported in **Table S5**.

Structure Calculation and refinement. Structure calculations were performed with the program CYANA 2.1 [17,18]. A total of 2000 random conformers were subjected to 65000 steps of a simulated annealing process. The 20 conformers with the lowest target function constituted the final family. Each member of the family was subsequently submitted to refinement in explicit solvent with the Amber-16 package [19]. The force field parameters for the 4Fe-4S cluster were taken as in similar systems [20]. A value of 50 kcal mol-1 $Ã^{-2}$ was used as force constant for the NOE and paramagnetic NMR restraints whereas a values of 32 kcal $mol^{-1}rad^{-2}$ was used for torsion angle restraints. The quality of the structure was evaluated in terms of deviations from ideal bond lengths and bond angles and through Ramachandran plots obtained using the programs with PSVS 1.5 program [21].

**Table S5**. Summary of conformationally-restricting constraints and structure quality factors of the AMBER energy minimized family of conformers of PioC

| | **PioC[a] (20 Conformers)** |
|---|---|
| **Total number of meaningful NOE upper distance constraints[b] :** | 344 |
| Intra-residue | 136 |
| Inter-residue | |
| Sequential (\|i-j\| = 1) | 103 |
| Medium-range (\|i-j\| < 4) | 56 |
| Long-range (\|i-j > 5) | 49 |
| **Total meaningful dihedral angle restraints :** | 51 |
| Phi | 26 |
| Psi | 25 |
| **Total number of paramagnetic NMR Restraints:** | 189 |
| Upper Distances constraints derived from $R_{1,2\ para}$ | 175 |
| Cys $\beta CH_2$ dihedral angle constraints $\chi_2$ | 4 |
| 1D NOEs between Cys bCH2 and neighboring residues | 7 |
| H-bonds linking HN donor atoms to S$\gamma$ of the preceding cluster-bound Cys | 3 |
| **Cluster** | |
| Geometrical parameters defining the cluster (upl) | 26 |
| Geometrical parameters defining the cluster (lol) | 26 |
| | |
| **Residual NOEs constraint violations[c] :** | |
| **Distance violations / structure** | |
| Between 0.1 -0.2 Å | 3.2 |
| Between 0.2 -0.5 Å | 0.3 |
| > 0.5 Å | 0 |
| RMS of Distances violations per meaningful distance constraint (Å): | 0.03 |
| Maximum distance violation [d] | 0.33 Å |
| **Residual PRE violations [e]** | |
| **Distance violations / structure** | |
| Between 0.1 -0.2 Å | 7.6 |
| Between 0.2 -0.5 Å | 2.0 |
| > 0.5 Å | 0 |
| RMS of Distances violations per meaningful distance constraint (Å): | 0.05 |
| **Dihedral angle violations / structure:** | |
| 1-10 ° | 3.3 |
| > 10° | 0 |
| **RMS violations per meaningful dihedral angle constraints (°):** | 0.82 ° |
| Maximum dihedral angle violation [d] | 6.90 ° |
| **Average RMSD to the mean (Å):** | |
| Residue range 5-50 (backbone atoms) | 0.62±0.11Å |
| Residue range 5-50 (all heavy atoms) | 1.14±0.13Å |
| residual CYANA Target Function ($Å^2$) | 1.92±0.13 |
| **Structure Quality Factors - overall statistics:** | **Z-score[g]** |
| Procheck G-factor e (phi / psi only) [f] | -3.30 |
| Procheck G-factor e (all dihedral angles) [f] | -4.91 |
| Verify3D | -4.01 |
| ProsaII (-ve) | 1.24 |
| MolProbity clashscore | 1.29 |
| **Ramachandran Plot Summary from Procheck [f]** | |
| Most favoured regions | 74.2% |
| Additionally allowed regions | 20.9% |
| Generously allowed regions | 3.40% |
| Disallowed regions | 1.40% |
| **Ramachandran Plot Statistics from Richardson's lab [f]** | |
| Most favoured regions | 84.6% |
| Allowed regions | 12.4% |
| Disallowed regions | 3.0% |

[a] The data are calculated over the 20 conformers representing the NMR structure. The mean value and the standard deviation are given

[b] Number of meaningful constraints for each class.

[c] Analyzed for residues 1 to 54, The analysis has been performed with PSVS 1.5 program considering the 344 meaningful NOEs

[d] Largest distance or dihedral angle constraint violation among all the reported structures

[e] Analyzed for residues 1 to 54, The analysis has been performed with PSVS 1.5 program considering the 175 Upper Distances constraints derived from $R_{1,2\ para}$

[f] Selected residue ranges: 5-50

[g] With respect to mean and standard deviation for a set of 252 X-ray structures < 500 residues, of resolution <= 1.80 Å, R-factor <= 0.25 and R-free <= 0.28; a positive value indicates a 'better' score. Z-score generated using PSVS 1.5

Comparison among different family of structures. Backbone and all heavy atoms RMSD obtained for each family of structure are summarized in **Table S6**. As discussed in text, it appears that: i) the highest precision is obtained when all available structural constraints are used, ii) the NOE-only and the PRE-only families of structures have very similar precision; iii) the contribution of the constraints arising from the cluster-bound residues, i.e. involving the first coordination sphere of the cluster, is extremely important. Indeed, the addition of 14 constraints (less than 5% of the total number of diamagnetic NOES) provides a 20% improvement in both backbone and all atoms RMSD (see the comparison between the first two columns of **Table S6**.

**Table S6**

| **Constraints used in Structure Calculation** | *NOE-only** | *NOE-only*** | *full-set* | *PRE-only* |
|---|---|---|---|---|
| Backbone RMSD[$] (residues 5-50) | 1.04 ±0.29 Å | 1.27±0.19 Å | 0.62±0.11Å | 1.31±0.27Å |
| All heavy at. RMSD[$] (residues 5-50) | 1.81±0.30 Å | 1.95±0.22 Å | 1.14±0.13Å | 2.00±0.32Å |

[$] *Data are related to the family of structure obtained upon CYANA calculation followed by 30 ps (15 000 steps with a time step of 2.0 fs) of restrained molecular dynamics at constant temperature and constant pressure (1.0 bar) using AMBER 16;*

*NOE-only** 344 meaningful NOEs + 26 geometrical parameters of the cluster + 14 constraints arising from cluster-bound residues (4 $\chi_2$ dihedral angle constraints -7 NOEs + 3 Hbonds)

*NOE-only*** 344 meaningful NOEs + 26 geometrical parameters of the cluster without constraints arising from cluster bound residues.

*full-set:* 344 meaningful NOEs + 26 geometrical parameters of the cluster + 14 constraints arising from cluster-bound residues +178 PREs

*PRE-only:* 26 geometrical parameters of the cluster + 14 constraints arising from cluster-bound residues +178 PREs

## 6. $^1$H $R_1$ and $R_2$ relaxation measurements and PRE constraints

Relaxation rates $^1$H $R_1$ and $^1$H $R_2$ were measured using a 11.7 T Bruker AVANCE 500 equipped with a triple resonance, inverse detection, cryoprobe (TXI) or a 14.0 T Bruker AVANCE NEO 600, equipped with a room temperature triple resonance inverse detection probe. For the $H_N$ $R_1$ rates, two series of experiments were used to measure longitudinal relaxation rates. For slow relaxing signals, a standard $^{15}$N-HSQC was edited with a non selective $^1$H inversion recovery building block. Fifteen experiments were collected (See Table S1), using a recycle delay of 4 s and inversion recovery delays of 20 ms, 30 ms, 40 ms, 50 ms, 60 ms, 80 ms, 100 ms, 120 ms, 160 ms, 200 ms, 300 ms, 400 ms, 600 ms, 800 ms, 1s. In order to measure relaxation rates of signals severely affected by the hyperfine relaxation, $R_1$ H

$_N$ rates were measured also with an IR-$^{15}$N-HSQC-AP experiment [22]. Fourteen experiments were collected using a recycle delay of 150 ms, an INEPT transfer delay (formally 1/(4J) ) of 710 us and inversion recovery delays of 2.0 ms, 4.0 ms, 6.0 ms, 10 ms, 15 ms, 20 ms, 25 ms, 30 ms, 40 ms, 50 ms, 60 ms, 80 ms, 120 ms, 200 ms. In both series of experiments, the intensities of the $^{15}$N HSQC spectra fitted according to three parameter fitting $I(t)= I(0)*[1-2\exp(-t*R1)] +C$ .

Transverse relaxation rate $^1H_N$-$R_2$, were also measured with two different approaches. For signals relatively far from the paramagnetic center, relaxation rates were measured using an experiment where a variable delay is inserted during the INEPT transfer of a $^{15}$N HSQC experiment [23]. Fourteen experiments were recorded, using a 4s recycle delay, a 1200 us selective $^1H_N$ inversion pulse for $^3JH_NH_\alpha$ decoupling, relaxation delays of 8.0 ms, 12 ms, 16 ms, 28 ms, 40 ms, 52 ms, 64 ms, 76 ms, 88 ms, 112 ms, 136 ms, 160 ms, 200 ms and 240 ms. To measure $^1$H $R_2$ rates of signals that are strongly affected by the hyperfine interaction, a new experiment, termed $R_2$-weighted $^{15}$N-HSQC-AP has been developed, as reported in **Figure S5**. $^1H_N$-$R_2$ measurements were obtained from a series of sixteen $R_2$-weighted $^{15}$N-HSQC-AP experiments recorded using recycle delays of 150 ms and INEPT transfer periods of 0.1 ms, 0.2 ms, 0.3 ms, 0.4 ms, 0.6ms, 0.8 ms, 1.0 ms, 1.2 ms, 1.4 ms, 1.6 ms, 2.0 ms, 2.4 ms, 2.8 ms, 3.2 ms, 4.0 ms, 5.0 ms. All relaxation data were analyzed using the Bruker Topspin Dynamics Center.

The $R_1$ and $R_2$ rates obtained with the different methods were compared and, for each residue, the value with the lower standard deviation in the fitting was considered. As expected, slow relaxing $^1$H resonances were better fitted using an in-phase $^{15}$N HSQC as editing spectrum and experimental conditions typical of diamagnetic systems, i.e. a long recycle delay, and longer relaxation periods. Conversely, fast relaxing signals were better fitted using the IR-HSQC-AP and the $R_2$ weighted HSQC-AP sequences. We found that all signals having $R_1$ values faster than 30 $s^{-1}$ (13 out of 48 total $R_1$ measurements) and $R_2$ values faster than 45 $s^{-1}$ (18 out of 49 total $R_2$ measurements) were better fitted using the tailored sequences. The results are summarized in **Table S7**.

For $^1H$ signals observed in $^{13}C$ HSQC experiments, $R_1$ $^1H$ rates were measured by inserting a non-selective $^1H$ inversion recovery filter prior to a standard $^{13}C$ HSQC experiments and fitting the intensities of the $^{13}C$ HSQC spectra as described above. Seventeen experiments were collected using a 2s relaxation delay and inversion recovery delays of 1.0 ms, 5.0 ms, 10 ms, 20 ms, 30 ms, 40 ms, 60 ms, 70 ms, 90 ms, 120 ms, 200 ms, 300 ms, 400 ms, 600 ms, 800 ms, 1.2 s, 2.0 s. The results are also reported in **Table S7**. $^1H$ relaxation rates of Cysteines H$\beta$ and H$\alpha$ protons were measured from an IR- $^{13}C$ HSQC-AP experiment, shown in **Figure S6**. The experiments were performed using acquisition and recycle delays of 17 ms and 65 ms, respectively. An INEPT transfer dealy of 600 us was used throughout the series. Ten experiments were performed using inversion recovery delays of 500 us, 2.0 ms, 4.0 ms, 6.0 ms, 8.0 ms, 10 ms, 15 ms, 30 ms, 50 ms, 80 ms.

<u>Conversion of Relaxation rates into distance constraints</u>. Each set of relaxation rates , i.e. $R_1$ $H_N$, $R_2$ $H_N$, $R_1$ H$\alpha$, $R_1$ $^1H$ side chains, $R_1$ $^1H$ Cysteine resonances, were analyzed independently. For each set of data, the diamagnetic contribution to the observed rate was estimated by taking the average value of those residues that are not affected by the hyperfine interaction. For this analysis were taken into consideration only those residues that, according to the internal dynamics measured with $^{15}N$ relaxation shown **Figure S7**, do not exhibit local internal motions. The $R_{1,2\ para}$ contribution is then calculated according to

$R_{1obs}= R_{1dia} + R_{1para}$.

Then, each set of $R_{1,2\ para}$ values is converted into a distance (d) according to an equation of the form $d=(A/R_{1,2})^{1/6}$, where the constant A was calibrated empirically. The calculation of the A factor from the Solomon equation describing the dipolar coupling between electron spin and nuclear spin (ref, Bertini libro) can not be reliably performed for a number of reasons: i) the electronic correlation time describing the electron relaxation of the $Fe^{+2.5}$ ions is dependent on the magnetic coupling operative within the cluster; therefore the effective $\tau_e$ is expected to be significantly different from what observed in isolated, high spin Fe(III) and Fe(II) ions. Its value can only be estimated based on $R_1$ and $R_2$ values of atoms at fixed distance from the cluster and from previously studied HiPIPs. ii) according to the metal centered approximation, which is mandatory for the Solomon equation (ref x), the unpaired electron spin density should be considered as fully localized onto the metal ions. In the case of a cluster, this is clearly an oversimplification of the problem, because the electron spin is also partly delocalized among all cluster atoms. Additionally, it is also known that the unpaired sin density is partly delocalized also onto the ligands, via both spin delocalization and spin polarization mechanisms. iii) paramagnetism arises from the population of the excited states of the electron spin energy ladder, which in turn depends from the antiferromagnetic coupling among the iron ions of the cluster (ref), which cannot be safely predicted. Provided all the above consideration, the use of an empirical calibration coefficient is recommended.

The conversion from $R_{1,2}$ values to distance has been previously done made according to an equation of the form:

$R_{1,2para} = A/(\Sigma d_i^6)$ (1) where di are the distances from the nuclear spin to the four iron ions.

Eq(1) accounts for atoms being at similar distances from two different Fe ions; however eq(1) still assumes that the electron spin density is fully localized onto the Iron ions. Therefore, to properly convert $R_{1,2\ para}$ values into upl we prefer to replace the $\Sigma d_i^6$ term with the power sixth distance from the center of mass of the cluster. In order to perform this, a special linker made of 100 pseudo-residues called LL2 was added at the end of the protein sequence. The "atoms" of these linker "residues" have zero mass and zero Van-der-Waals radii, thus the linker can freely pass through the structure during simulated annealing. The last residue of the linker is an ION residue (cyana library) which has been subsequently linked at fixed distances with the four Iron and with the four sulfur ions of the cluster, with van der Waals contact taken to zero in order to avoid distortions or additional contribution to the overall energy. For the AMBER refinement, the linker and the ION residue have been removed and, for each PRE constraint, the center of mass of the cubane has been replaced with the closest iron ion of the cluster (or with the two closest iron ions when ambiguous metal-to-proton distances occur), and the upper limit distance reduced accordingly.

| **Table S7** | | | | | | |
|---|---|---|---|---|---|---|
| **n protein** | | **HN $R_2$** | **std** | **HN $R_1$** | **std** | |
| **VAL** | **1** | | | | | |
| **THR** | **2** | **11,1** | 0,4 | **12,9** | 0,5 | |
| **LYS** | **3** | **15,4** | 1,3 | **15,9** | 0,8 | |
| **LYS** | **4** | **19,1** | 1,5 | **3,83** | 0,1 | |
| **ALA** | **5** | **23,4** | 0,8 | **5,41** | 0,17 | |
| **SER** | **6** | **17,8** | 0,6 | **2,21** | 0,08 | |
| **HIS** | **7** | **21,3** | 0,8 | **4,90** | 0,08 | |
| **LYS** | **8** | **32,8** | 2,2 | **14,8** | 0,5 | |
| **ASP** | **9** | **16,9** | 0,7 | **2,74** | 0,13 | |
| **ALA** | **10** | **19,6** | 0,8 | **7,28** | 0,25 | |
| **GLY** | **11** | **20,0** | 1,4 | **4,77** | 0,12 | |
| TYR | 12 | **29,9** | 1,5 | **11,1** | 0,5 | |
| GLN | 13 | **53,2** | 4,1 | **16,6** | 0,6 | |
| GLU | 14 | **19,4** | 1,1 | **3,18** | 0,11 | |
| SER | 15 | **21,6** | 1,0 | **3,19** | 0,18 | |
| PRO | 16 | | | | | |
| ASN | 17 | **22,8** | 1,1 | **4,69** | 0,19 | |
| GLY | 18 | **25,1** | 1,5 | **7,50** | 0,36 | |
| ALA | 19 | **44,4** | 2,8 | **22,5** | 1,3 | |
| LYS | 20 | **17,8** | 0,8 | **2,79** | 0,19 | |
| ARG | 21 | **44,4** | 3,4 | **12,0** | 0,6 | |
| CYS | 22 | **311** | 18 | **116** | 32 | |
| GLY | 23 | **60,4** | 6,8 | **19,0** | 0,6 | |
| THR | 24 | **42,0** | 2,8 | **14,6** | 0,3 | |
| CYS | 25 | **60,1** | 2,8 | **37,6** | 3,8 | |
| ARG | 26 | **43,6** | 3,6 | **32,0** | 8,5 | |
| GLN | 27 | **246** | 18 | **160** | 13 | |
| PHE | 28 | **146** | 9 | **89,1** | 5,6 | |
| ARG | 29 | **49,3** | 3,0 | **25,0** | 1,0 | |
| PRO | 30 | | | | | |
| PRO | 31 | | | | | |
| SER | 32 | **25,1** | 1,5 | **7,50** | 0,36 | |
| SER | 33 | **24,2** | 0,8 | **5,75** | 0,27 | |
| CYS | 34 | **42,5** | 4,0 | **17,6** | 0,5 | |
| ILE | 35 | **81,4** | 5,7 | **48,0** | 5,1 | |
| THR | 36 | **107** | 9 | **100** | 7 | |
| VAL | 37 | **156** | 16 | **129** | 12 | |
| GLU | 38 | **24,2** | 0,8 | **8,71** | 0,17 | |
| SER | 39 | **19,7** | 1,0 | **3,89** | 0,10 | |
| PRO | 40 | | | | | |
| ILE | 41 | **27,7** | 1,3 | **5,61** | 0,12 | |
| SER | 42 | **42,8** | 2,4 | **13,9** | 0,4 | |
| GLU | 43 | **19,2** | 0,5 | **5,18** | 0,06 | |
| ASN | 44 | **27,4** | 1,5 | **5,62** | 0,11 | |
| GLY | 45 | **59,3** | 4,3 | **24,7** | 1,2 | |
| TRP | 46 | **69,5** | 5,2 | **42,7** | 2,3 | |
| CYS | 47 | **194** | 10 | **134** | 29 | |
| ARG | 48 | **53,5** | 4,2 | **35,4** | 2,1 | |
| LEU | 49 | **167** | 15 | **129** | 7 | |
| TYR | 50 | **47,7** | 7,8 | **39,8** | 10,5 | |
| ALA | 51 | **240** | 36 | | | |
| GLY | 52 | **23,0** | 1,3 | **6,38** | 0,19 | |
| LYS | 53 | **18,0** | 0,9 | **6,42** | 0,18 | |
| ALA | 54 | **7,7** | 0,3 | **2,15** | 0,07 | |
| Trp sc | 46 | **33,9** | 2,1 | | | |
| | | diamag-netic ex-periments | | para-magnetic experi-ments | | |

| | | $R_1$ Hα | $R_1$ Hβ | $R_1$ Hγ | $R_1$ Hδ | $R_1$ Hε | R other |
|---|---|---|---|---|---|---|---|
| VAL | 1 | | | | | | |
| THR | 2 | 2,50 | 2,66 | 2,73 | | | |
| LYS | 3 | 2,34 | 3,60 | 2,80 | 2,56 | 2,14 | |
| LYS | 4 | 3,79 | 3,93/4,41 | 4,40/6,36 | | 3,94/4,25 | |
| ALA | 5 | 2,93 | 7,20 | | | | |
| SER | 6 | 2,92 | 2,23/2,47 | | | | |
| HIS | 7 | 10,2 | 4,06/4,13 | | | | |
| LYS | 8 | 2,39 | 3,16/3,36 | 3,17/3,15 | 2,77/2,69 | 2,44 | |
| ASP | 9 | 1,96 | 2,72/2,80 | | | | |
| ALA | 10 | 8,16 | 5,58 | | | | |
| GLY | 11 | 5,01/3,09 | | | | | |
| TYR | 12 | 16,7 | 24,6/15,8 | | 56,7 | | |
| GLN | 13 | 3,22 | 2,78/3,41 | 5,03/5,09 | | | |
| GLU | 14 | 4,79 | 3,00/2,79 | 4,02 | | | |
| SER | 15 | 2,30 | 2,94/2,96 | | | | |
| PRO | 16 | 4,99 | 4,04/3,61 | 3,10/6,02 | 2,46/2,58 | | |
| ASN | 17 | 2,24 | 4,48/4,23 | | | | |
| GLY | 18 | 1,37/2,14 | | | | | |
| ALA | 19 | 1,55 | 2,31 | | | | |
| LYS | 20 | 5,78 | 6,90/8,35 | 4,60 | 3,67/4,27 | 2,35/2,32 | |
| ARG | 21 | 29,8 | 8,09/3,03 | 5,26/7,89 | 4,08/4,01 | | |
| CYS | 22 | 53,6 | 368/130 | | | | $^{13}$Cβ 67 |
| GLY | 23 | 7,81/6,15 | | | | | |
| THR | 24 | | 4,47 | 3,77 | | | |
| CYS | 25 | 59,2 | 313/282 | | | | $^{13}$Cβ 59<br>$^{13}$Cα 13 |
| ARG | 26 | 10,2 | 8,02/4,87 | 8,65/6,67 | 3,65/4,54 | | |
| GLN | 27 | 29,6 | 1,61 | 1,76/2,03 | | | |
| PHE | 28 | 2,34 | 45,1 | | 44,5 | | |
| ARG | 29 | 4,94 | 5,32/6,11 | 5,96/5,40 | 4,22/4,39 | | |
| PRO | 30 | 3,36 | 2,43/2,50 | 2,31/2,48 | 3,28/4,38 | | |
| PRO | 31 | 2,15 | 1,64/1,97 | 4,79/2,02 | 1,77/1,42 | | |
| SER | 32 | 4,67 | 3,45/3,44 | | | | |
| SER | 33 | | 5,89/5,72 | | | | |
| CYS | 34 | 69,3 | 164/588 | | | | $^{13}$Cβ 65 |
| ILE | 35 | 8,32 | 10,9 | 6,84/10,0/7,56 | 4,95 | | |
| THR | 36 | 3,00 | 2,49 | 2,98 | | | |
| VAL | 37 | 21,6 | | 59,9 | | | |
| GLU | 38 | 4,41 | 4,50/4,53 | 3,66/3,58 | | | |
| SER | 39 | 6,34 | 3,08/2,69 | | | | |
| PRO | 40 | 3,62 | 2,12/2,79 | 4,69/4,24 | 1,88/1,87 | | |
| ILE | 41 | 4,42 | 34,1 | 31,9/37,5/95,0 | 2,71 | | |
| SER | 42 | 5,17 | 5,72/8,04 | | | | |
| GLU | 43 | 16,2 | 4,91/4,19 | 4,03/3,26 | | | |
| ASN | 44 | 5,53 | 3,35/3,62 | | | | |
| GLY | 45 | 14,9/10,9 | | | | | |
| TRP | 46 | | 16,7/22,7 | | 36,4 | 84,3 | |
| CYS | 47 | | 130/450 | | | | $^{13}$Cβ 67 |
| ARG | 48 | 13,8 | 4,23 | 7,25/6,63 | 3,10/4,39 | | |
| LEU | 49 | 12,7 | | 4,44 | 5,37 | | |
| TYR | 50 | 2,70 | | | 97,0 | | |
| ALA | 51 | 10,7 | 13,1 | | | | |
| GLY | 52 | 4,12/7,56 | | | | | |
| LYS | 53 | 3,64 | 4,31 | 4,75 | 3,55 | | |
| ALA | 54 | 2,56 | 2,08 | | | | |
| | | diamagnetic exp. | | paramagnetic exp. | | | |

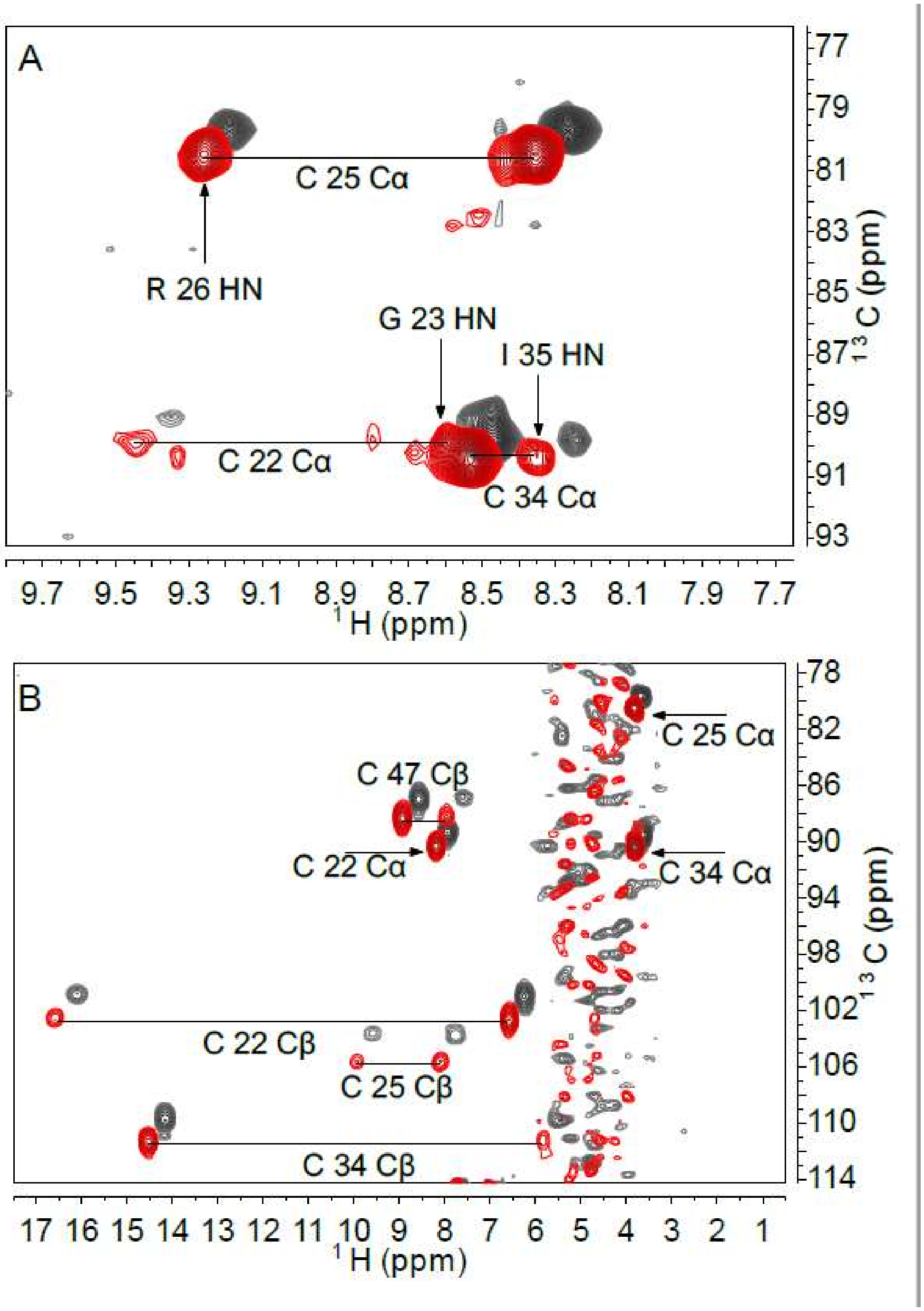

**Figure S1.** 500 MHz, H(N)CA (**A**) and $^{13}$C HSQC (**B**) experiments at 288K (black) and 298K (red) optimized for peaks involving fast relaxing resonances. Peculiar 13C shift values allows the sequential assignment of three out of the four cluster-bound Cysteines in the HNCA, while the $CH_2$ pairs are unambiguously identified via $^{13}$C HSQC.

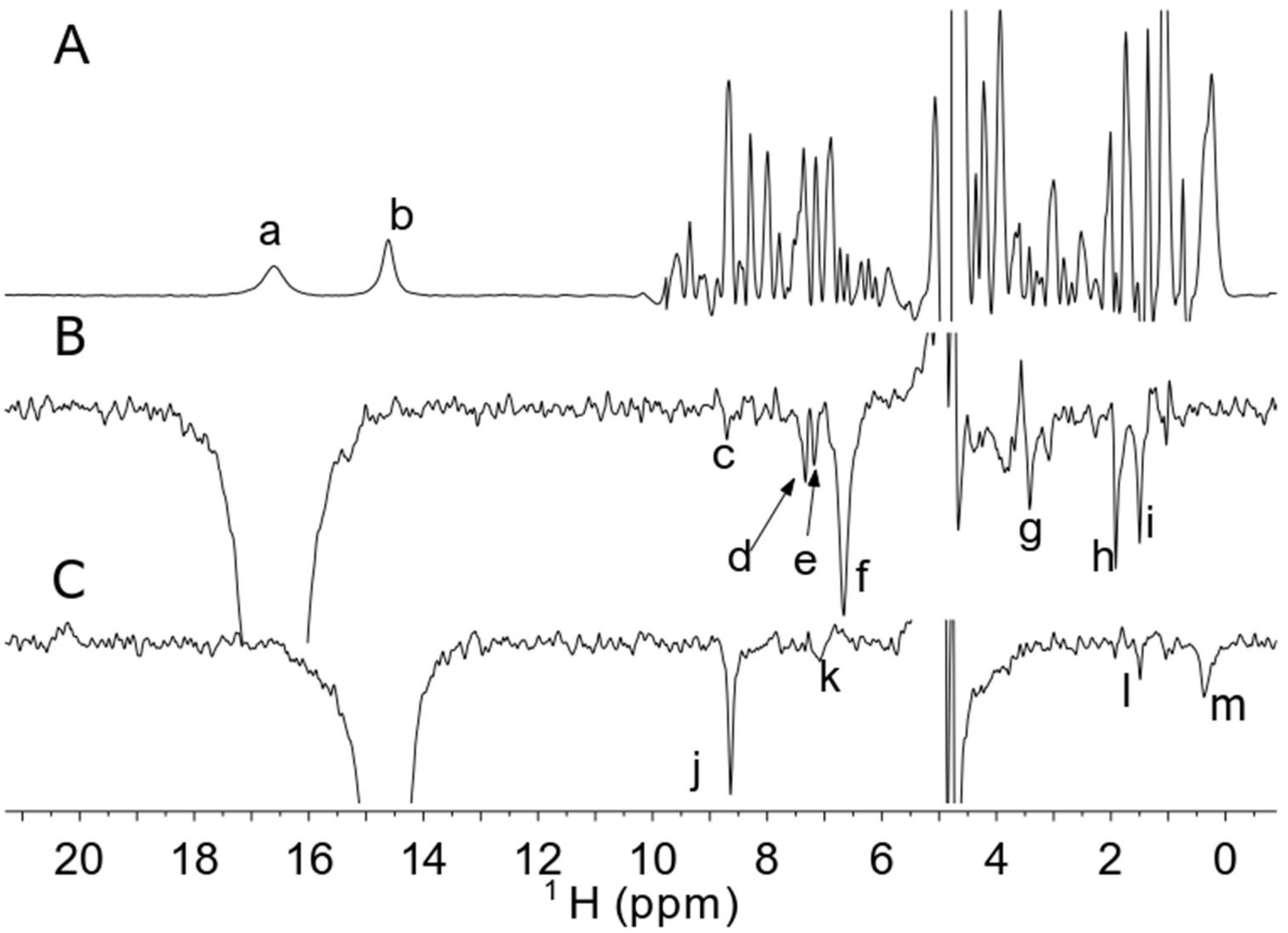

**Figure S2**. **A** 1D $^1$H NMR spectrum of PioC, optimized to observe hyperfine shifted and fast relaxing resonances. Two isolated peaks (labeled a-b) are observed. **B-C** 1D NOE difference spectra obtained upon selective saturation of signals a-b. The NOE peaks observed in the difference spectra, labeled c-m, are used to perform the assignment of signals a and b. Experiment were performed at 400 MHz and 298K. About 400.000 scans were acquired for each NOE, using the inversion recovery sequence experiments and collecting the difference between on and off resonance selective irradiation

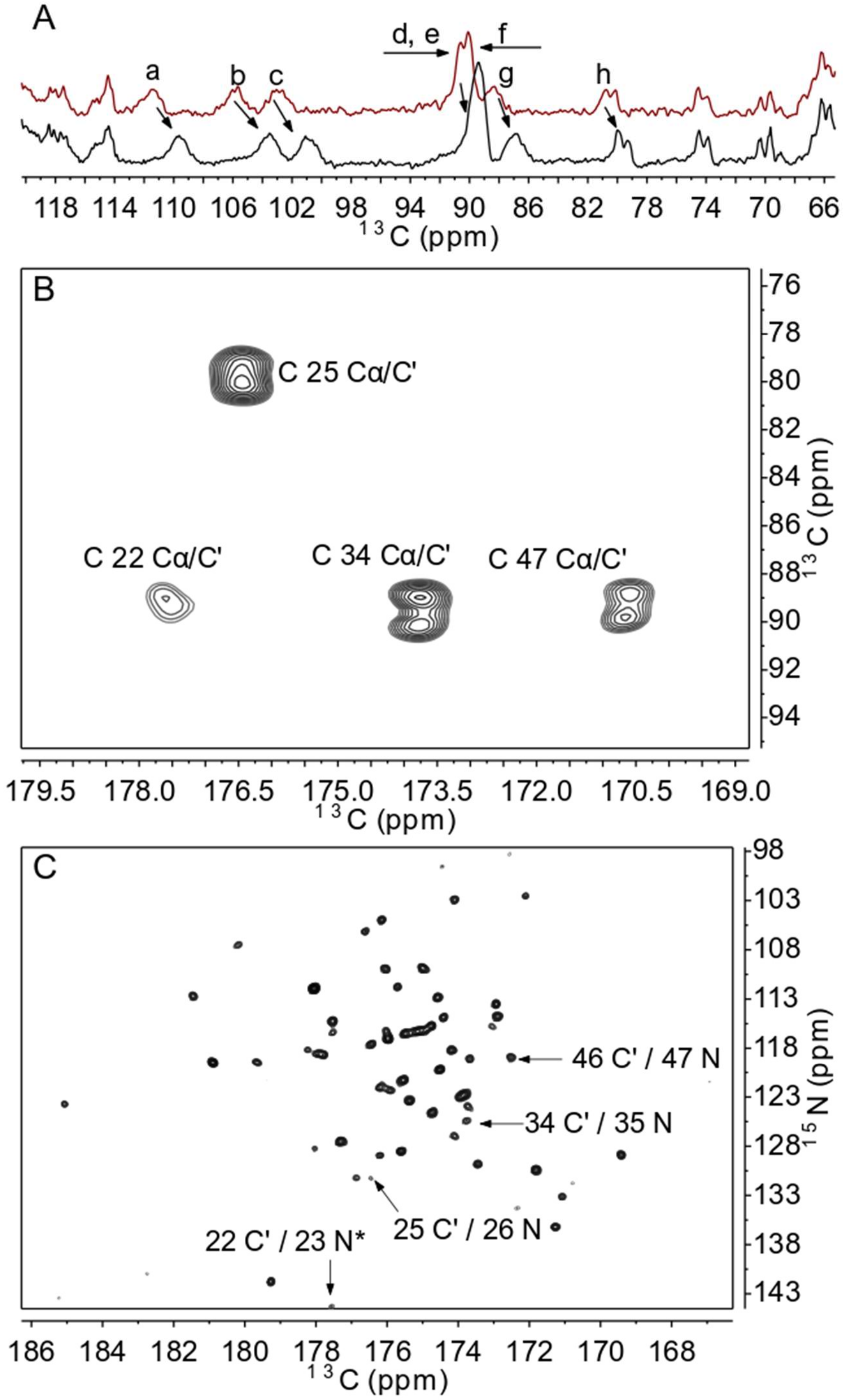

**Figure S3**. **A** 1D $^{13}$C NMR spectra of PioC, optimized to observe fast relaxing resonances. Figure shows the spectral region where we expect to observe $^{13}$C□ and $^{13}$C□ of cluster-bond residues. Spectrum in black is recorded at 288 K, in red at 298K. **B** $^{13}$C-$^{13}$C COSY spectrum (upper diagonal part) showing the connectivities between C' and C□ signals. Acquisition and processing parameters are optimized to identify connectivities among fast relaxing resonances. In both dimensions no $^{1}$H decoupling has been used. **C** $^{13}$C-$^{1r}$N CON experiment. Signals involving C' spins from cluster-bound Cysteines are labeled in figure. All the above experiments were performed at 175 MHz, using $^{13}$C direct detection

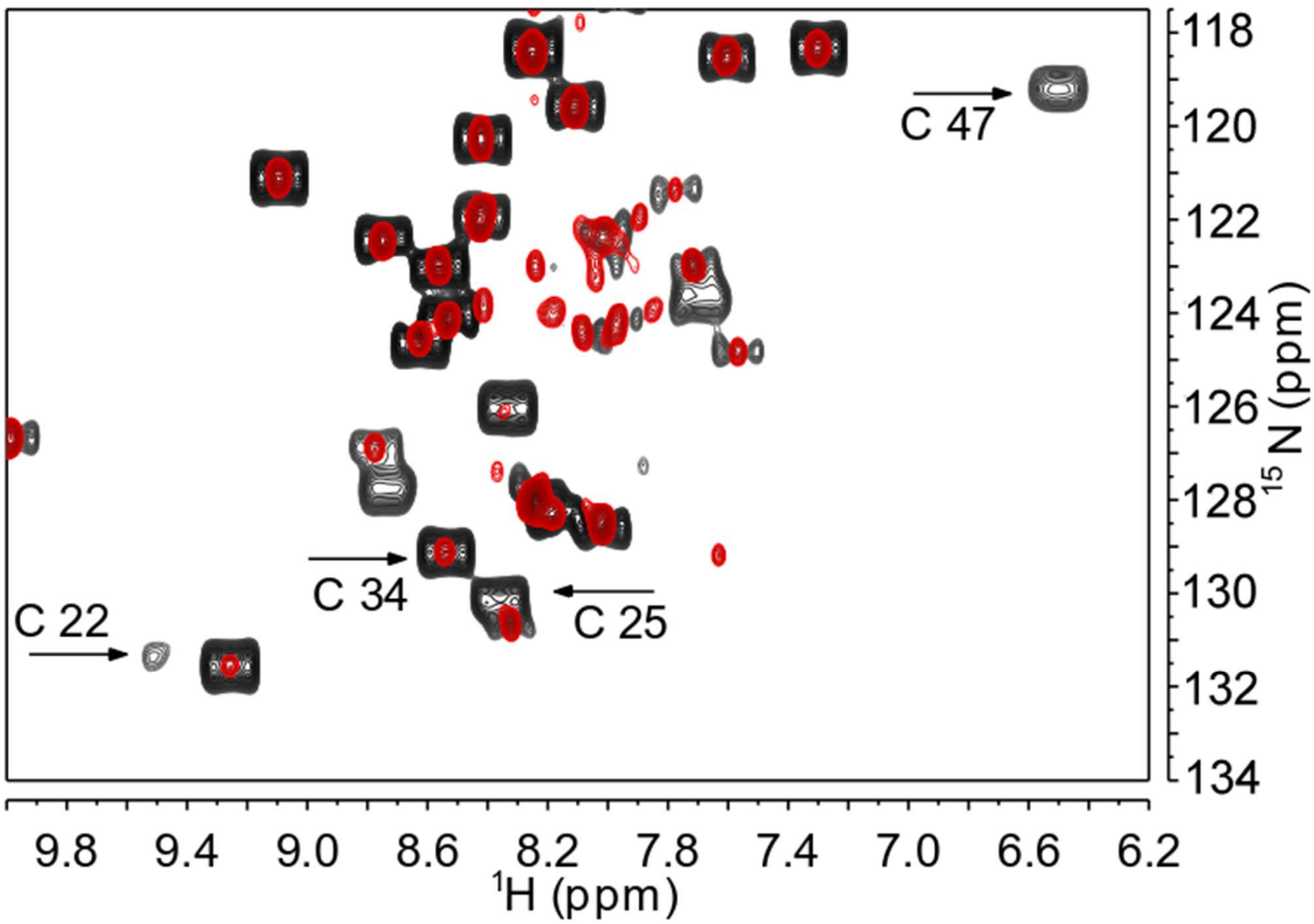

**Figure S4.** Expanded region of a $^{15}$N HSQC-AP experiment where the four cluster-bound cysteines are labeled. Spetrum has been recorded at 500 MH, 298K, and a standard in-phase HSQC experiment is overlaid in red for comparison purposes.

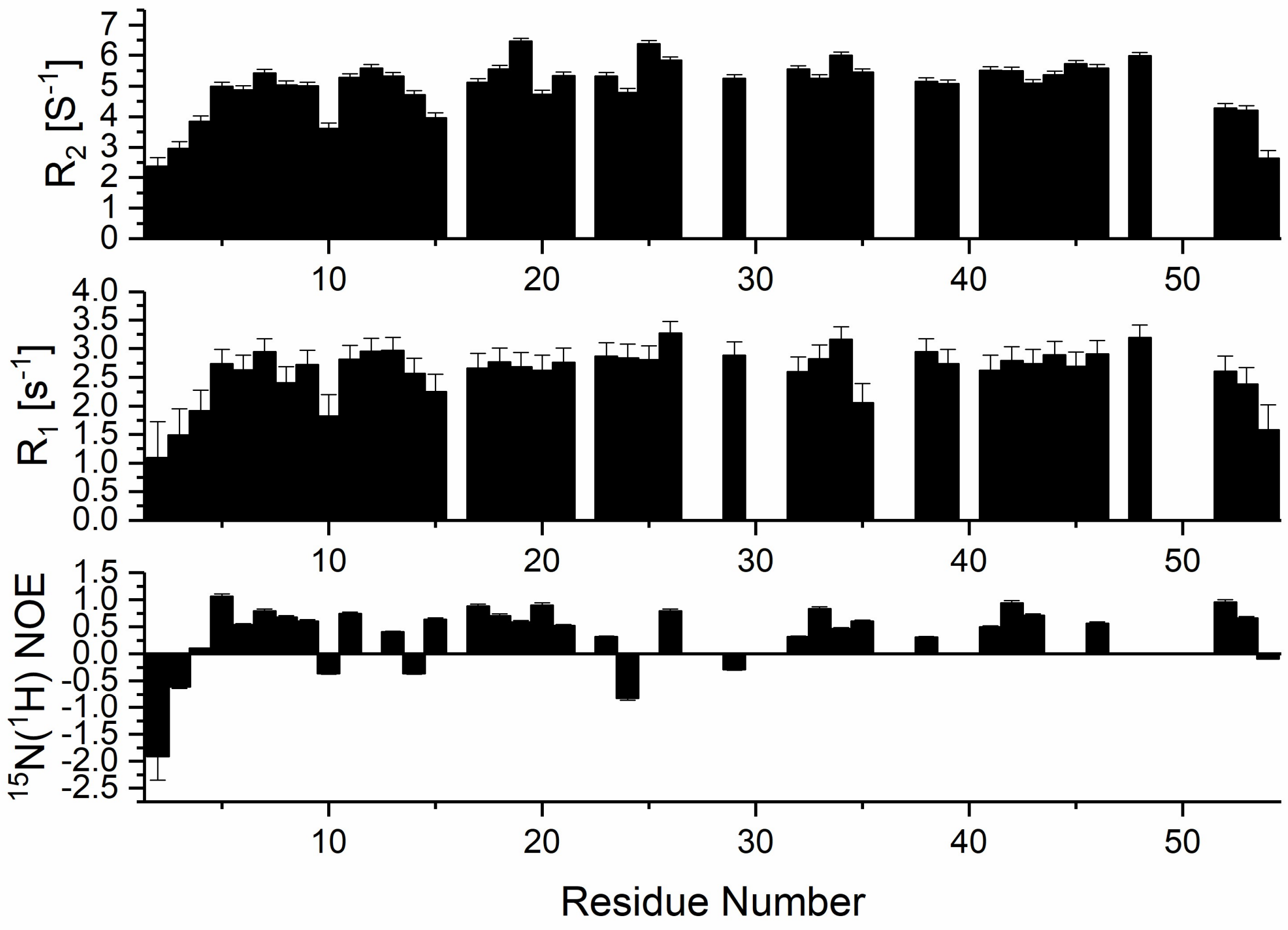

**Figure S5.** Experimental $^{15}N$ $R_1$, $R_2$ rates and heteronuclear NOEs as obtained from the $^{15}N$ relaxation experiments performed at 500 MHz, 298K on a $^{15}N$ labelled PioC sample

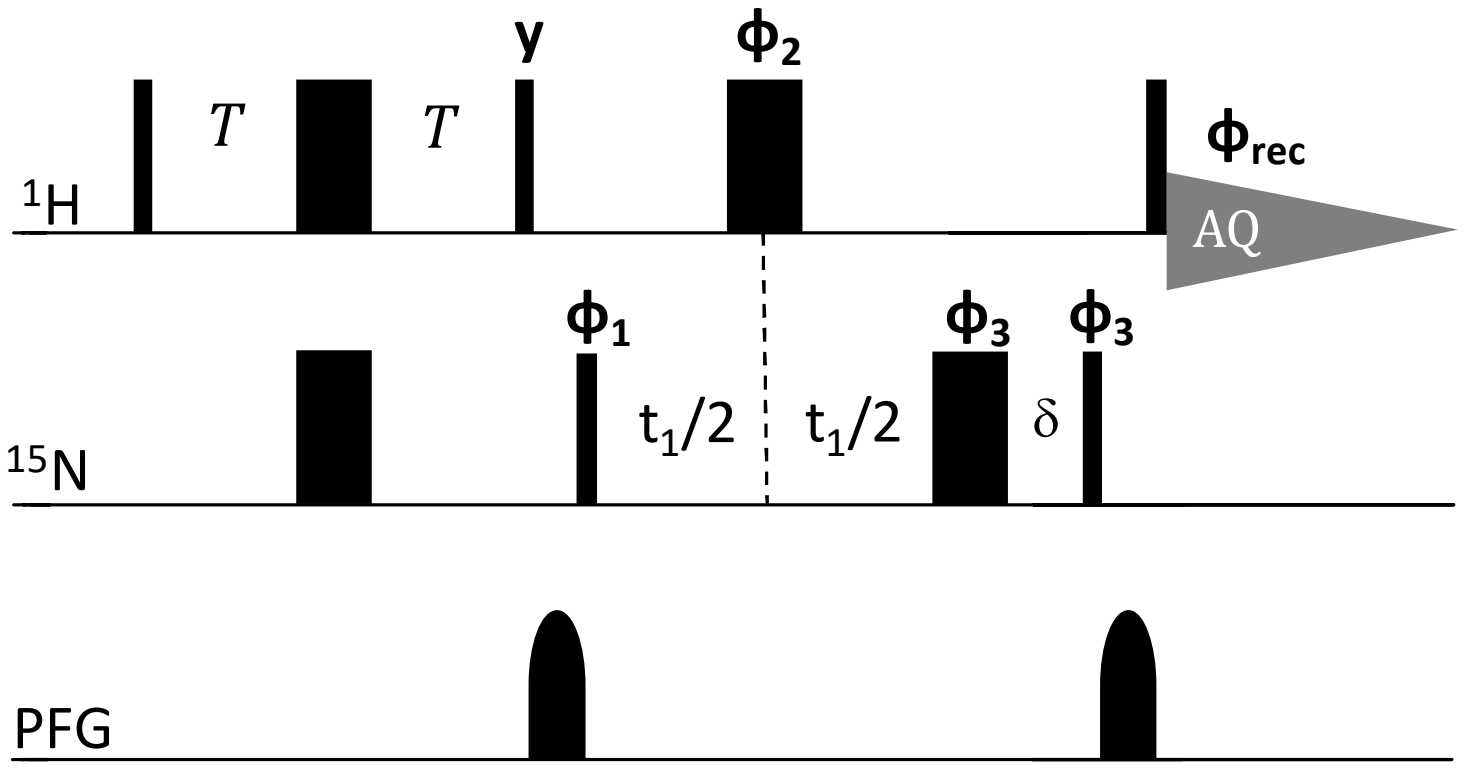

$\phi_1$ = x, -x
$\phi_2$ = 2 (x), 2 (-x)
$\phi_3$ = 4 (x), 4 (-x)
$\phi_{rec}$ = x, -x, x, -x, -x, x, -x, x

**Figure S6**. Pulse sequence used for the measurement of $^1H_N$ $R_2$ relaxation rates. The experiment is an $^{15}$N-HSQC experiments with detection in antiphase. For each HN peak, the intensity of the doublet as a function of *T* provides the values of $^1$H transverse relaxation rate

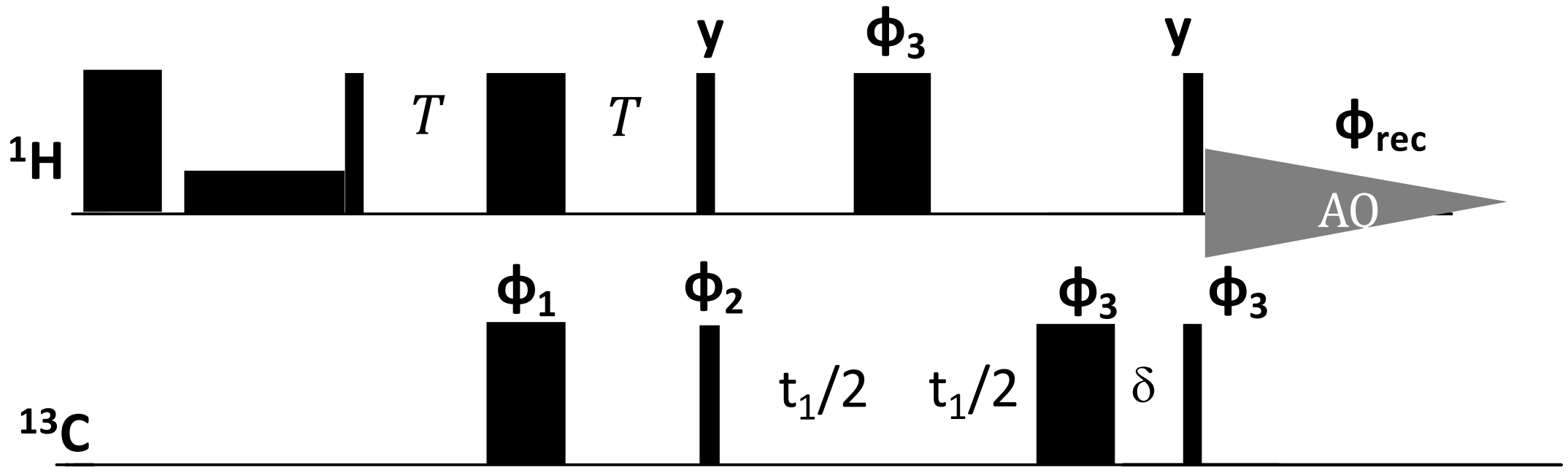

$\phi_1$ = 8 (x), 8(-x)
$\phi_2$ = x, -x
$\phi_3$ = 2 (x), 2 (-x)
$\phi_{rec}$ = x, -x, x, -x, -x, x, -x, x

**Figure S7.** Pulse sequence used for the measurement of $^1H_C$ $R_1$ relaxation rates. The experiment is a $^{13}$C-IR-HSQC experiment with detection in antiphase. For each HC peak, the intensity of the doublet as a function of the inversion recovery delay provides the values of $^1$H longitudinal relaxation rates